\documentclass[10pt,final,journal,letterpaper,twoside,twocolumn]{IEEEtran}

\usepackage{cite}
\usepackage{graphicx,color,epsfig,rotating}
\usepackage{amsfonts,amsmath,amsthm,amssymb,bbm,dsfont}
\usepackage{algorithm,algorithmic}
\usepackage{subfigure}

\newtheorem{thm}{Theorem}

\newtheorem{lem}{Lemma}
\newtheorem{cor}{Corollary}

\newtheorem{defi}{Definition}
\newtheorem{rem}{Remark}

\newcounter{storeeqcounter}
\newcounter{tempeqcounter}

\def\argmin{\mathop{\rm argmin}}

\makeatletter
\newcommand{\vast}{\bBigg@{3.5}}
\newcommand{\Vast}{\bBigg@{5}}
\makeatother

\def\pK{{\pmb K}}\def\p0{{\pmb 0}}

\def\snr{{\small\textsf{SNR}}}

\def\cn{{\mathcal{CN}}}

\newfont{\bbc}{msbm10 scaled 1100}

\newcommand{\Cc}{{\cal C}}

\newcommand{\Lc}{{\cal L}}

\newcommand{\Oc}{{\cal O}}

\newcommand{\Sc}{{\cal S}}

\def\argmin{\mathop{\rm argmin}}

\def\snr{{\small\textsf{SNR}}}

\IEEEoverridecommandlockouts

\begin{document}

\title{{On the High-SNR Capacity of the Gaussian Interference Channel and New Capacity Bounds }}

\author{Junyoung Nam,  ̃\IEEEmembership{Member, ̃IEEE}
\thanks{This work was supported by Institute for Information \& communications Technology Promotion (IITP) grant funded by the Korea government (MSIP) [R0101-16-244]. 
The material in this paper was presented in part at the IEEE International Symposium on Information Theory (ISIT),  Cambridge, Massachusetts, Jul./Aug. 2012.}
\thanks{J. Nam is with the department of Wireless Communications and Networks, Fraunhofer Heinrich Hertz Institute (HHI), 10587 Berlin, Germany (e-mail: junyoung.nam@hhi.fraunhofer.de). He was with the Wireless Communication Division, Electronics and Telecommunications Research Institute (ETRI), Daejeon, Korea. } 
\thanks{Copyright (c) 2016 IEEE. Personal use of this material is permitted.  However, permission to use this material for any other purposes must be obtained from the IEEE by sending a request to pubs-permissions@ieee.org.}
}

\maketitle

\begin{abstract}
The best outer bound on the capacity region of the two-user Gaussian Interference Channel (GIC) is known to be the intersection of regions of various bounds including genie-aided outer bounds, in which a genie provides noisy input signals to the intended receiver. The Han and Kobayashi (HK) scheme provides the best known inner bound. {The rate difference between the best known lower and upper bounds on the sum capacity remains as large as 1 bit per channel use especially around $g^2=P^{-1/3}$, where $P$ is the symmetric power constraint and $g$ is the symmetric real cross-channel coefficient. In this paper, we pay attention to the \emph{moderate interference regime} where $g^2\in (\max(0.086, P^{-1/3}),1)$.} We propose a new upper-bounding technique that utilizes noisy observation of interfering signals as {genie signals} {and applies time sharing to the genie signals at the receivers.} {A conditional version of the worst additive noise lemma is also introduced to derive new capacity bounds. The resulting upper (outer) bounds on the sum capacity (capacity region) are shown to be tighter than the existing bounds in a certain range of the moderate interference regime. Using the new upper bounds and the HK lower bound, we show that $R_\text{sym}^*=\frac{1}{2}\log \big(|g|P+|g|^{-1}(P+1)\big)$ characterizes the capacity of the symmetric real GIC to within $0.104$ bit per channel use in the moderate interference regime at any signal-to-noise ratio (SNR). We further establish a high-SNR characterization of the symmetric real GIC, where the proposed upper bound is at most $0.1$ bit far from a certain HK achievable scheme with Gaussian signaling and time sharing for $g^2\in (0,1]$. In particular, $R_\text{sym}^*$ is achievable at high SNR by the proposed HK scheme and turns out to be the high-SNR capacity at least at $g^2=0.25, 0.5$.} {It is finally pointed out that there are two \emph{effective} subregimes at high SNR in the weak interference regime, where $g^2\in(0,1]$}. 
\end{abstract}


\section{Introduction}
\label{sec:intro}


{Finding the capacity of the interference channel is a long-standing open problem in network information theory \cite{Car78,EK12}. Such a capacity region would reveal optimal ways of managing interference, which is one of the most fundamental and challenging issues in current wireless networks tending to be more interference-limited.}
In the past decade, the Gaussian Interference Channel (GIC) has received great attention as a primitive model for a wireless network with mutually interfering links. Nevertheless, the capacity of even the two-user GIC has not been fully characterized yet. {The capacity region is known} in some cases such as the very strong interference regime \cite{Car75} and the strong interference regime \cite{Sat81,Han81}. 
In \cite{Sat78}, Sato showed that the capacity region of the degraded GIC is outer-bounded by a certain degraded broadcast channel whose capacity region is fully known. In \cite{Cos85}, Costa proved that the Sato outer bound can be used for the one-sided GIC (or Z interference channel) owing to the equivalence between the one-sided GIC and the degraded GIC, which in turn can be an outer bound for the general (two-sided) GIC. Sato \cite{Sat77} derived another outer bound by allowing the receivers to cooperate, based on the fact that the capacity of GIC depends only on the marginal noise distributions so that correlation among Gaussian noises does not affect the capacity. {Carleial \cite{Car83} developed another outer bound} by decreasing the noise power. For inner bounds, the best known achievable region is given by Han and Kobayashi (HK) \cite{Han81}, which uses rate splitting, time sharing, and simultaneous {non-unique} decoding of the intended signal and part of the interfering signal. The full HK achievable region is known to be formidable  to compute due to  numerous degrees of freedom in computation such as the cardinality of the time-sharing parameter. 
{In \cite{Cho08}, the HK achievable region was compactly expressed after Fourier-Motzkin elimination and an upper bound on the cardinality of the time-sharing parameter was known to be at most 8 for the general discrete memoryless interference channel. For the GIC with Gaussian signals, the cardinality of the time-sharing parameter can be less than 4 \cite{Mot09} and Sason \cite{Sas04} proposed a particular HK scheme easily computable with the cardinality of 4.}

A significant progress to the characterization of the capacity region of the two-user GIC in the weak interference regime, where the interference is weaker than the intended signal, was initiated by Kramer \cite{Kra04} and Etkin, Tse, and Wang \cite{Etk08}. Inspired by the classical approaches given by Sato \cite{Sat77,Sat78}, Careial \cite{Car83}, and Costa \cite{Cos85}, {Kramer derived two tighter outer bounds by improving upon those bounds in \cite{Sat77,Sat78,Car83,Cos85} for the GIC.} Etkin \emph{et al.} elaborated on the design of a  genie signal that provides some noisy observation of the intended signals to the receivers and derived an outer bound, referred to as Etkin-Tse-Wang (ETW) outer bound, tighter than the Kramer bound in a regime of very weak interference. By designing a more general genie signal, the subsequent works in \cite{Mot09,Sha09,Ann09} independently achieved an improvement on the ETW bound, which will be called the ``enhanced ETW" upper bound in this paper. More importantly, {they proved that treating an interference from an independent and identically distributed (i.i.d.) Gaussian codebook as noise is optimal in the noisy interference regime}, where cross-channel coefficients are very weak and SNR should not be too high, thereby implying that any sophisticated interference management scheme does not increase the capacity in that regime. The enhanced ETW upper bound on the sum capacity was further tightened later by \cite{Etk09, Cha11}. Therefore, the best capacity outer bound for the two-user GIC in the weak interference regime is known as the intersection of regions given by the Kramer bound\cite{Kra04} and the {genie-aided} 
bounds in \cite{Etk08,Mot09,Sha09,Ann09,Etk09,Cha11}.

{This long-standing open problem in network information theory has very recently attracted renewed attention. In particular, \cite{Cos16} established the slope of the HK region with Gaussian signaling at the corner point of the Gaussian Z-interference channel. In \cite{Bei16}, the authors established how far one can go with Gaussian inputs in the single-letter HK region and proved that capacity region exhibits a discontinuity of slope around the sum-rate point for a subset of the very weak interference channel.
The optimality of the well-known ``Costa's corner points" was proved by \cite{Pol16}. In \cite{Liu15}, the authors developed a new version of the genie-aided sum-rate outer bound, which recovers all known capacity results. It is interesting to notice that treating interference from a non-Gaussian codebook as noise without time sharing was shown in \cite{Dyt16} to be optimal to within a constant gap. The mixed non-Gaussian codebook is the sum of a discrete (acts as a common message) and a Gaussian (as a private message) random variable similar to rate splitting in the HK scheme.} 

{It follows from the prior works that the sum capacity of the two-user GIC remains unknown  in the \emph{moderate interference} regime, where $g^2\in (\max(0.086, P^{-1/3}),1)$ for the symmetric case. In particular, the rate gap between the existing upper bounds and lower bounds is still as large as $1$ bit {at high SNR, which we call ``missing" one bit in this work.} Note that this moderate interference becomes a bottleneck in interference networks and hence is of particular interest because strong interference can be rather cancelled out at the receivers and treating interference as noise is optimal in noisy interference. 
As a consequence, how much one can reduce the rate gap in moderate interference and hence finding the ``missing" one bit are an intriguing open problem in interference management.} 

{In this paper, we accordingly focus on the moderate interference regime and derive new capacity upper (outer) bounds for the two-user GIC that are tighter than the known bounds in a large portion of the regime of interest. 
To this end, we propose a new genie-aided approach. Notice that the well-known Etkin-type genie-aided approach in \cite{Etk08,Mot09,Sha09,Ann09,Etk09,Cha11} provides the receivers with a noisy version of their own intended signals. In the proposed approach, a genie instead provides the receivers with a noisy version of their interfering signals.
The role of the proposed genie signal is to replace arbitrarily distributed interfering random sequences with i.i.d. Gaussian random sequences in order to cast the original interference channel into a mathematically more tractable one. In order to overcome a major difficulty in upper-bounding some negative entropy terms pointed out earlier by \cite{Tel07}, we further introduce a conditional version of the worst additive noise lemma \cite{Dig01}. {Another key idea is to jointly use the Etkin-type and the proposed genie signal and to apply \emph{time sharing} to each of them}. 
}


{A main contribution of this work is as follows: Using the HK lower bound and our upper bounds, we  show that 
\begin{align} \label{eq:intro-1}
    R_\text{sym}^* =\frac{1}{2}\log \big(|g|P+|g|^{-1}(P+1)\big)
\end{align} 
characterizes the capacity of the symmetric real GIC to within $\frac{1}{2}\log\frac{2}{\sqrt{3}}\; (\approx 0.104)$ bit per channel use  in the moderate interference regime for any SNR. While the proposed upper bounds are lower-bounded by $R_\text{sym}^*$, a particular HK scheme with Gaussian signals and  with the cardinality of time-sharing parameter of $2$ (inspired by Sason \cite{Sas04}) is shown to be optimal to within $0.125$ bit in the moderate interference regime for $P\ge 23.3$. At high SNR, it is further shown that the particular HK scheme achieves $R_\text{sym}^*$. Moreover, $R_\text{sym}^*$ turns out to be the high-SNR capacity at least at $g^2=0.25, 0.5$. {As $P\rightarrow\infty$, the HK scheme is shown to be at most $0.1$ bit far from the proposed upper bound for the entire weak interference regime of interest, i.e., $g^2\in (0, 1]$.} {Accordingly, the well-known one bit gap between the best known lower and upper bounds is almost filled. Moreover, we show that there are two effective subregimes in the weak interference regime at high SNR \emph{in the scale of} $g^2$.}} 

Another implication of the main results is that if the interference level is moderate, then any sophisticated interference management scheme including the HK scheme would not achieve a significant performance gain. 
This is because the rate gap between the orthogonal-resource (e.g., time/frequency division multiplexing (TDM/FDM)) approach referred to as the TDM scheme and $R_\text{sym}^*$ is at most $0.3995$ bit.  
This is somewhat counter-intuitive to the important insight provided by the {generalized degrees of freedom} \cite{Etk08}  implying that we could obtain a considerable potential gain over the TDM lower bound {around $g^2=P^{-1/3}$ (i.e., $\alpha=2/3$), where rate splitting becomes important. 
We also observe that the well-known ``W" shape in \cite[Fig. 11]{Etk08} is not translated well into the behavior of sum-rate bounds around $\alpha=2/3$, unless SNR is very high (say, $P \gg 1000$).}
It is further shown that our capacity outer bounds are tighter than the existing outer bounds for a certain range of channel parameters. 
{The proposed upper-bounding technique in this paper has been extended to better understand the general $K$-user GIC in the companion work \cite{Nam15b}.}


\begin{figure}
\hspace{-5mm} \center
  \includegraphics[scale=1]{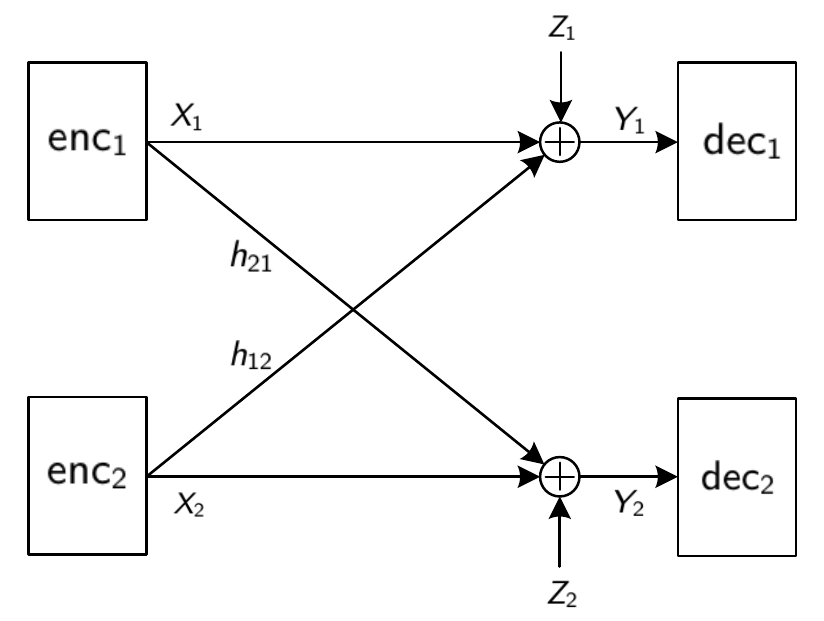}
  \caption{Two-user Gaussian interference channel. }\label{fig-a}
\end{figure}

The remainder of this paper is organized as follows. Section \ref{sec:IC} describes the channel model of the
two-user GIC that we study. In Section \ref{sec:BT}, we review the Etkin-type genie-aided upper-bounding technique  and introduce the new bounding technique along with a new worst additive noise lemma. In Sections \ref{sec:NB}, we derive new upper bounds on the sum capacity of the two-user GIC. Section \ref{sec:NB-A} is devoted to simplify the proposed upper bounds and characterize the sum capacity of the symmetric real GIC in the moderate interference regime. Section \ref{sec:HS} provides a high-SNR characterization of the symmetric capacity to within a constant gap.  In Section \ref{sec:OB}, we develop new outer bounds on the capacity region. We conclude this work in Section  \ref{sec:Con}.

\emph{Notations:} We use $X$ for a random variable and $X^n$ for a random sequence. Also, $\sigma^2_X$ denotes the variance of $X$. For a set $A$, we use $|A|$ to denote the cardinality of $A$. For $c\in \mathbb{C}$, let  $\mathfrak{R}\{c\}$ denote the real part of $c$. We denote $\cn(0,1)$ for circularly symmetric Gaussian random variables of unit variance.

\section{Two-User Gaussian Interference Channel}
\label{sec:IC}

The standard two-user complex GIC can be written as 
\begin{align} \label{eq:A-1}
  Y_1&=X_1+h_{12}X_2+Z_1 \nonumber \\ Y_2&=h_{21}X_1+X_2+Z_2
\end{align}
where the inputs $X_1 \in \mathbb{C}$ and $X_2 \in \mathbb{C}$ have the average power constraints of $P_1$ and $P_2$, respectively, $h_{ij} \in \mathbb{C}$ is the (static) cross-channel coefficient between transmitter $j$ and receiver $i$, and $Z_1$ and $Z_2$ are distributed as $\cn(0,1)$ and independent of inputs.

Let $M_1$ and $M_2$ be independent, uniformly distributed messages over $[1:2^{nR_1}]$ and $[1:2^{nR_2}]$, and let $X^n_1 \in \boldsymbol{\mathcal{X}}_1, X^n_2 \in \boldsymbol{\mathcal{X}}_2, Y^n_1 \in \boldsymbol{\mathcal{Y}}_1, Y^n_2 \in \boldsymbol{\mathcal{Y}}_2$ be the random sequences induced by encoders $\mathrm{enc}_i: [1:2^{nR_i}]\rightarrow \boldsymbol{\mathcal{X}}_i, i=1,2,$ and the channel, respectively, where the channel input $X_i^n$ satisfies the average power constraints of $P_i$ such that $||X_i^n||^2\le nP_i, i=1,2.$ {Denoting the decoding functions at the receivers as $\mathrm{dec}_i: \boldsymbol{\mathcal{Y}}_i \rightarrow[1:2^{nR_i}] , i=1,2,$ respectively, the average probability of decoding error for user $i$ is defined as $P_{e,i}^{(n)} =\mathbb{E} [ \Pr (\mathrm{dec}_i(\boldsymbol{\mathcal{Y}}_i)\neq m_i )]$. 
A rate pair $(R_1,R_2)$￼ is said to be achievable if there exists a sequence of $(2^{nR_1},2^{nR_2})$ codes with $\lim_{n\rightarrow \infty} P_{e,i}^{(n)}=0, i=1,2$.} The capacity region ￼of the two-user GIC is defined as the closure of the set of all achievable rate pairs {$(R_1,R_2)$}. 
Throughout this paper, let $X_{iG} \sim \mathcal{CN}(0,P_i)$ (or $\sim\mathcal{N}(0,P_i)$ for the real GIC) and $Y_{iG}=X_{iG}+Z_i$ for all $i$ and the corresponding i.i.d. random sequences are denoted by $X_{iG}^n$ and $Y_{iG}^n$, respectively. Fig. \ref{fig-a} depicts the above two-user Gaussian interference channel.

We also consider the \emph{symmetric real} two-user GIC for simplicity. The {symmetric real} GIC with constant channel coefficients is given by
\begin{align} \label{eq:A-1b}
  Y_1&=X_1+g X_2+Z_1 \nonumber \\
  Y_2&=X_2+ gX_1+Z_2
\end{align}
where $g\in \mathbb{R}$ is the common real cross-channel coefficient and $P_1=P_2=P$. 

{As mentioned in the introduction, we focus on a particular regime of cross-channel coefficients $h_{ij}$ in GIC, where the interference level is moderate  as follows:  
\begin{defi} \label{def-1} \normalfont
The moderate interference regime of GIC is a subregime of the well-known weak interference regime, where $|h_{12}|^2$ and $|h_{21}|^2$ are less than one, such that 
\begin{align} 
  |h_{12}|^2&\in \big(\max(0.086, P_1^{-1/3}),1\big) \nonumber \\
  |h_{21}|^2&\in \big(\max(0.086, P_2^{-1/3}),1\big). \nonumber
\end{align}
In the symmetric real GIC, we have $$g^2\in (\max(0.086, P^{-1/3}),1).$$
\end{defi}

For the symmetric real case, the second element in $\max(\cdot,\cdot)$ follows from the generalized degrees of freedom \cite{Etk08} such that $$\alpha\triangleq\frac{\log g^2P}{\log P}=\frac{2}{3} \ \Longleftrightarrow \ g^2=P^{-1/3}. $$
The first element $0.086$ ($\approx(1-\sqrt{1/2})^2$) will be discussed later in Sec. \ref{sec:NB-A}.}

\section{Upper-Bounding Approaches}
\label{sec:BT}

In this section, we briefly review the well-known genie-aided bounding approach along with the resulting genie-aided outer bounds and  introduce a new upper-bounding approach. These two bounding approaches are the main building blocks to be separately or jointly used throughout this paper.

\subsection{{Etkin-type Genie-aided Approach}} 

{A genie-aided bounding technique implies that a genie provides some side information to either one or both receivers. 
In this work, the \emph{Etkin-Type} genie signals $S_i^n$ is defined as the side information that contains at least the input $X_i^n$ such that $I(X^n_i;Y^n_i,S^n_i)\neq I(X^n_i;Y^n_i|S^n_i)$.} To the best of our knowledge, this holds for all genie signals so far identified useful. One may further restrict the genie signals $S_i^n$ such that they are conditionally i.i.d. Gaussian sequences given the corresponding input sequence $X_i^n$ as in  \cite{Etk08,Tel07,Mot09,Sha09,Ann09}, which we assume in this work as well.\footnote{A more general genie signal can be found in \cite{Kra04,Etk09}.}

The key idea of the work by Etkin, Tse, and Wang \cite{Etk08} is to devise a genie in a sophisticated manner, yielding a tighter outer bound on the capacity region than the bounds in \cite{Kra04} for a certain range of channel coefficients.
The authors designed the genie signal such that
{\begin{align} \label{eq:A-2}
  S_1&=h_{21}X_1+N_1' \nonumber \\     
  S_2&=h_{12}X_2+N_2'
\end{align}
where $N_i'\sim\cn(0,1), i=1,2,$ is independent of everything else}, so as to obtain the following useful property:
\begin{align} \label{eq:A-3}
  h(Y^n_1|X^n_1,S^n_1)&=h(S^n_2) \nonumber \\ h(Y^n_2|X^n_2,S^n_2)&=h(S^n_1)
\end{align} 
where $X_i^n$ denotes the sequence of length $n$ of the input $X_i$ and $S_i$ denotes side information to receiver $i$ for $i=1,2$. Using the genie signal, the authors provided a meaningful outer bound in \cite[Thm. 3]{Etk08} and showed that 
{the gap between a simplified HK region and the ETW outer bound is at most 1 bit per channel use, irrespective of the channel gains and power constraints.}
The resulting genie-aided outer bounds on the capacity region of the GIC defined by (\ref{eq:A-1}) in the weak interference regime (i.e., {$|h_{12}|\le 1$ and $|h_{21}|\le 1$}) is as follows.

\begin{thm}[ETW bound \cite{Etk08}] \label{lem-1} \normalfont
The capacity region of the two-user GIC in the weak interference regime is contained in the set of rate pairs $(R_1, R_2)$ satisfying
\begin{subequations}\label{eq:A-7}
\begin{align} 
  R_1 &\le I(X_{1G};Y_{1G}|X_{2G}) \label{eq:A-7a} \\ 
  R_2 &\le I(X_{2G};Y_{2G}|X_{1G}) \label{eq:A-7b} \\ 
  R_1+R_2 &\le I(X_{1G};Y_{1G}|X_{2G}) +I(X_{2G};Y_{2G}) \label{eq:A-7c} \\ 
  R_1+R_2 &\le I(X_{1G};Y_{1G}) +I(X_{2G};Y_{2G}|X_{1G}) \label{eq:A-7d} \\ 
  R_1+R_2 &\le I({X}_{1G};{Y}_{1G},S_{1G}) +I({X}_{2G};{Y}_{2G},S_{2G}) \label{eq:A-7e} \\ 
  2R_1+R_2 &\le I(X_{1G};Y_{1G}|X_{2G})+I(X_{1G};Y_{1G}) \nonumber \\ &\ \ \ +I({X}_{2G};{Y}_{2G},S_{2G}) \label{eq:A-7f} \\ 
  R_1+2R_2 &\le I({X}_{1G};{Y}_{1G},S_{1G})+I(X_{2G};Y_{2G}|X_{1G}) \nonumber \\ &\ \ \ +I(X_{2G};Y_{2G}). \label{eq:A-7g}
\end{align}
\end{subequations}
\end{thm}

The ETW outer bound was improved by choosing genie signals more elaborately in the subsequent works \cite{Mot09,Sha09,Ann09}, which used the following natural generalization of the genie signal in (\ref{eq:A-2}):
\begin{align} \label{eq:A-9}
  S_1&=h_{21}X_1+N_1   \nonumber \\     
  S_2&=h_{12}X_2+N_2  
\end{align}
{where $N_i \sim\cn(0,\sigma_{N_i}^2)$ is independent of $(X_1,X_2)$.} Here we allow $N_i$ to be correlated to $Z_i$ with correlation coefficient $\rho_{N_i}$, for $i=1,2$. The resulting outer bound, which we call the enhanced ETW bound, can be stated as follows.

\begin{thm}[Enhanced ETW bound \cite{Mot09,Ann09}] \label{thm-2a} \normalfont
The capacity region of the two-user GIC  in the weak interference regime is outer-bounded by the set of rate pairs $(R_1, R_2)$ satisfying (\ref{eq:A-7}) in which (\ref{eq:A-7e}), (\ref{eq:A-7f}), and (\ref{eq:A-7g}) are tightened by using the genie signal (\ref{eq:A-9}) and the extremal inequality in \cite{Mot09} (or the entropy power inequality (EPI) in \cite{Ann09}). 
\end{thm}

A common bounding technique using the genie signals in (\ref{eq:A-2}) and (\ref{eq:A-9}) for the ETW bound and its enhanced version is that we can derive 
\begin{align}
  n(R_1-\epsilon_n) &\le I(X^n_1;Y^n_1) \nonumber \\ 
  &\le I(X^n_1;Y^n_1,S_1^n) \nonumber \\ 
  &= h(S^n_1)-h(S^n_1|X^n_1) +h(Y^n_1|S^n_1) \nonumber \\ 
  &\hspace{5mm} -h(Y^n_1|X^n_1,S^n_1)   \label{eq:BT-1c} \\ 
  &\le h(S^n_1)-h(Y^n_1|X^n_1,S^n_1) +nh(Y_{1G}|S_{1G})\nonumber \\ 
  &\hspace{5mm} -nh(N_1)    \label{eq:BT-1}
 \end{align}
where $\epsilon_n\rightarrow 0$ as $n\rightarrow \infty$ and in the last inequality we used {the genie signals in (\ref{eq:A-9})} and the fact that Gaussian maximizes entropy. In order to further bound (\ref{eq:BT-1}), the ETW bound makes use of (\ref{eq:A-3}), whereas the enhanced ETW bound jointly considers $I(X^n_1;Y^n_1,S_1^n) +I(X^n_2;Y^n_2,S_2^n)$ and applies the worst additive noise lemma  \cite{Dig01} to $h(S^n_1)-h(Y^n_2|X^n_2,S^n_2)$ to obtain
\begin{align}
  h(S^n_1)-h(Y^n_2|X^n_2,S^n_2) &\le nh(S_{1G})-nh(h_{21}X_{1G}+Z_2|N_2) \nonumber \\
  &= nh(S_{1G})-nh(h_{21}X_{1G}+V_{N_2})   \label{eq:BT-1b}
 \end{align}
for $(N_1,N_2)$ satisfying the condition of $\sigma_{{N_1}}^2\le\sigma_{V_{N_2}}^2$, where $V_{N_2}^n$ is i.i.d. $\mathcal{N}(0,\sigma_{V_{N_2}}^2)$ with $\sigma_{V_{N_2}}^2=\sigma_{Z_2|N_2}^2$, and it does the same to $h(S^n_2)-h(Y^n_1|X^n_1,S^n_1)$.

The enhanced ETW bound can be further improved by designing a more general genie signal in \cite{Etk09}, which we call the ``further enhanced ETW bound" in this paper.

\subsection{{New Genie-aided Approach}}

In this section, we develop a new {genie-aided} upper-bounding approach. The key idea is to design {different genie signals  from (\ref{eq:A-9})} such that we can replace the arbitrary random sequence acting as interference signal to the intended receiver by a Gaussian random sequence whose components are i.i.d. in time.
Accordingly, this can be referred to as the \emph{change-of-interference} approach. 

We first present a special case of this new approach in the following. 
Given the channel inputs $X_1^n$ and $X_2^n$ in Section \ref{sec:IC}, we define the {genie signals} $(U_1, U_2)$ as
\begin{align} \label{eq:A-4b}
  U_1&=h_{12}X_2+W_1 \nonumber \\ U_2&=h_{21}X_1+W_2
\end{align}
where $W_i $ is a zero-mean circularly symmetric complex Gaussian random variable with variance $\sigma_{W_i}^2 \le 1$, correlated to $Z_i$ with correlation coefficient $\rho_{W_i}$ (i.e., $\mathbb{E}[Z_k W_k^*]=\rho_{W_k}\sigma_{W_k}$) and independent of everything else, for $i=1,2$. 
Using the corresponding change-of-interference sequence $U^n_1$ whose components are i.i.d. in time, we can upper-bound $R_1$ as 
\begin{align} 
    n(R_1-\epsilon_n) &\le I(X^n_1;Y^n_1) \nonumber \\ 
  &\overset{(a)}{\le} I(X^n_1;Y^n_1|U^n_1) \label{eq:BT-4a} \\
  &=h(Y^n_1|U^n_1)-h(Y^n_1|X^n_1,U^n_1) \nonumber \\
  &= h(Y^n_1|U^n_1)-h(h_{12}X^n_2+Z^n_1|h_{12}X^n_2+W^n_1) \nonumber \\
  &\overset{(b)}{=} h(Y^n_1|U^n_1)-h(Z^n_1-W^n_1|h_{12}X^n_2+W^n_1) \nonumber \\
   &=  h(Y^n_1|U^n_1) - h(h_{12}X^n_2+V^n_{W_1}) \nonumber \\ 
  &\hspace{5mm} +h(U^n_1) -nh(Z_1-W_1) \label{eq:BT-4} \\
   &=  h(X^n_1+Z_1^n-W_1^n|U^n_1) - h(h_{12}X^n_2+V^n_{W_1}) \nonumber \\
   &\ \ \ \ +h(U_1^n) -nh(Z_1-W_1) \label{eq:BT-5b}
 \end{align}
where $$V_{W_1}^n \text{ is i.i.d. } \mathcal{CN}(0,\sigma_{V_{W_1}}^2)$$
with $\sigma_{V_{W_1}}^2=\sigma_{W_1|Z_1-W_1}^2$. In the above inequalities, $(a)$ follows from the well-known inequality  between the conditional mutual information $I(X;Y|Z)$ and the unconditional mutual information $I(X;Y)$ \cite{Cov06}, as shown by 
\begin{align} \label{eq:BT-8}
   \text{If } p(x, y, z) &= p(x)p(z)p(y|x, z), \text{ then } \nonumber \\ 
  &\hspace{5mm} I(X;Y|Z) \ge I(X;Y).
\end{align}
Notice that we do not use the principle that Gaussian maximizes entropy such that $h(Y^n_1|U^n_1) \le nh(Y_{1G}|U_{1G})$, in contrast to (\ref{eq:BT-1}) in the Etkin-type approach. {In $(b)$, we rather changed the arbitrary random sequence $h_{12}X^n_2$ in the output $Y_1^n$ to the i.i.d. Gaussian random sequence $h_{12}W^n_1$ conditioned on $U_1^n$. This step makes the resulting multi-letter expressions more tractable.} It will be shown in Sections \ref{sec:NB} and \ref{sec:OB} that $h(X^n_1+Z_1^n-W_1^n|U^n_1)$ in (\ref{eq:BT-5b}) enables various upper bounds in conjunction with the conditional worst additive noise lemma in Lemma \ref{lem-4}. 

Likewise, we have
\begin{align} \label{eq:BT-4b}
  n(R_2-\epsilon_n) &\le h(Y^n_2|U^n_2) - h(h_{21}X^n_1+V^n_{W_2}) \nonumber \\ 
  &\hspace{5mm} +h(U^n_2) -nh(Z_2-W_2)
\end{align}
where $V_{W_2}^n$   is i.i.d. $\mathcal{CN}(0,\sigma_{V_{W_2}}^2)$ with $\sigma_{V_{W_2}}^2=\sigma_{W_2|Z_2-W_2}^2$.

\begin{rem} \normalfont
The change-of-interference approach applies first the Fano's inequality to the original channel and then identifies some useful auxiliary random sequences satisfying (\ref{eq:BT-8}). 
For the above special case, this approach can be cast into the genie-aided approach since $I(X^n_1;Y^n_1|U^n_1)=I(X^n_1;Y^n_1,U^n_1)$ due to the independence between $U_1^n$ and $X_1^n$, but not vice versa as mentioned earlier. 
{For the two-user GIC, the new genie-aided approach may be viewed as providing $S_1^n$ to $Y_2^n$ and $S_2^n$ to $Y_1^n$. The difference is that $N_1^n$ should be correlated with $Z_2^n $ rather than $Z_1^n $, and so $N_2^n$ be.} 
\end{rem}

We can develop a more general form of the change-of-interference approach, which is distinguished from the genie-aided approach. The details can be found in Appendix \ref{sec:app-0}. In this paper, however, we restrict our attention to the simple case in (\ref{eq:A-4a}).


\subsection{Conditional Worst Additive Noise Lemma}

{In order to derive our upper (outer) bounds, we need a new bounding tool} --- conditional version of the worst additive noise lemma \cite{Dig01} as follows: 


\begin{lem}[Conditional Worst Additive Noise Lemma] \label{lem-4} \normalfont
Let $X^n$ denote a random sequence with an average power constraint, $Z^n$ be i.i.d. $\mathcal{N}(0,\sigma_Z^2)$ independent of $X^n$ and let $U^n$ denote another random sequence with an average power constraint, correlated with $X^n$ but independent of $Z^n$. Suppose that the corresponding random vector sequence $(Z,X+Z,U)^n$ has an average covariance constraint such that $\frac{1}{n}\sum_{j=1}^n\mathrm{Cov}(Z_j,X_j+Z_j,U_j)\preceq \pK$. Also let $(Z,X_g+Z,U_g)$ be a zero-mean Gaussian random vector with covariance $\pK$. Then, we have
\begin{align} \label{eq:B-4}
  h(X^n|U^n)-&h(X^n+Z^n|U^n)\nonumber \\ &\le nh(X_g|U_g)-nh(X_g+Z|U_g)
\end{align}
where the equality holds if $X^n=X^n_g$ and $U^n=U^n_g$.
\end{lem}
\begin{IEEEproof} 
Refer to Appendix \ref{sec:app-1}.
\end{IEEEproof}

Even though the above proof is a natural generalization of \cite[Lem. 4]{Ann09}, Lemma \ref{lem-4} is not in terms of random \emph{vector} sequences and has a different covariance constraint. Meanwhile, it is somewhat analogous to the conditional extremal inequality in \cite[Thm. 8]{Liu07}. While the conditional extremal inequality is conditioned on a scalar random variable $U$ since its proof relies on the classical conditional EPI \cite{Ber74}, our inequality (\ref{eq:B-4}) is rather conditioned on the random sequence $U^n$. 
Using the conditional worst additive noise lemma, we can obtain the following useful lemma.

\begin{lem} \label{lem-8} \normalfont
Let $X^n$ and $Y^n$ denote arbitrary random sequences with average power constraints and let $Z^n$, $W^n$, $V^n$ be i.i.d. $\mathcal{N}(0,\sigma_Z^2)$, $\mathcal{N}(0,\sigma_W^2)$, $\mathcal{N}(0,\sigma_V^2)$, respectively, independent of both $X^n$ and $Y^n$. 
If 
\begin{align} \label{eq:B-24}
   \sigma_{V}^2\ge \sigma_{Z- W}^2
\end{align}
then we have
\begin{align} \label{eq:B-20}
   h(X^n&+Y^n+Z^n|Y^n+W^n) - h(X^n+V^n) \nonumber \\
  &\le nh(X_g+Y_g+Z|Y_g+W) \nonumber \\ 
  &\hspace{5mm} - nh(X_g+Y_g+Z+\tilde{V}|Y_g+W)
\end{align}
where {$\tilde{V} \sim \mathcal{N}(0,\sigma_{V}^{2}-\sigma_{Z- W}^2)$ is independent of $Z$ and $W$.}
\end{lem}

\begin{IEEEproof}
Refer to Appendix \ref{sec:app-2}.
\end{IEEEproof}

The above two lemmas are naturally applied to the complex-valued random variables and will be widely used to derive our various upper (outer) bounds in the rest of this paper.

\section{New Upper Bounds on the Sum Capacity}
\label{sec:NB}

Using the conditional worst additive noise lemma, we can combine the two main upper-bounding approaches in Sec.  \ref{sec:BT} in various ways to derive new upper bounds on the sum capacity  of the two-user GIC in the weak interference regime. Another key idea is to apply \emph{time sharing} to the Etkin-type and the proposed genie signal, respectively.

{
\subsection{Time Sharing on Side Information at the Receivers}
\label{sec:NB-0}

\begin{figure}
 \hspace{-3mm}
  \includegraphics[scale=1.05]{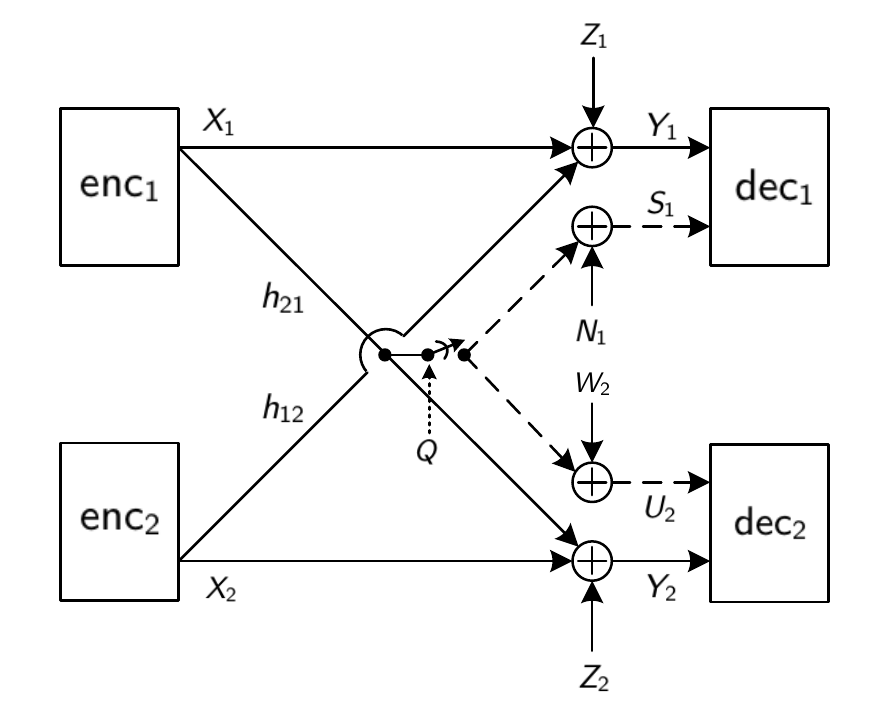}
  \caption{Genie-aided enhanced channel induced by \eqref{eq:ob-7d}. The two types of side information $S_1$ and $U_2$ are only active when the time-sharing parameter $Q$ equals $0$ with $\Pr(Q=0)=\frac{1}{2}$.}\label{fig-b}
\end{figure}

We first introduce a time-sharing operation with respect to side information at the receivers. 
Let $Q$ denote a time sharing random variable. With $|Q|=2$, we apply time sharing on the genie signals $S_{1}$ and $U_{2}$ as follow:  
\begin{subequations} \label{eq:IS-2}
\begin{align} \label{eq:IS-2a}
  \tilde{S}_{1}^n = \left\{ \begin{array}{ll}
  {S}_{1}^n &  \text{if } Q=0\\
  0 &  \text{if } Q=1
  \end{array} \right.
\end{align}
\begin{align} \label{eq:IS-2b}
  \tilde{U}_{2}^n = \left\{ \begin{array}{ll}
  U_{2}^n &  \text{if } Q=0\\
  0 &  \text{if } Q=1
  \end{array} \right.
\end{align}
\end{subequations}
where $S_{1}$ and $U_{2}$ are given by (\ref{eq:A-9}) and (\ref{eq:A-4b}), respectively. The random sequences $\tilde{S}_{1}^n$ and $\tilde{U}_{2}^n$ are conditionally independent given $Q$. 
Using Fano's inequality and letting $\mathrm{Pr}(Q=0)=\mathrm{Pr}(Q=1)=1/2$, we can write 
\begin{align} \label{eq:ob-7c}
  n(R_1+R_2&-2\epsilon_n) \nonumber \\
  &\le I({X}_{1}^n;{Y}_{1}^n) +I(X_{2}^n;Y_{2}^n) \nonumber \\
  &\le I({X}_{1}^n;{Y}_{1}^n,\tilde{U}_{1}^n) +I(X_{2}^n;Y_{2}^n,\tilde{U}_{2}^n)  \nonumber \\
  &\overset{(a)}{\le} I({X}_{1}^n;{Y}_{1}^n,\tilde{U}_{1}^n|Q) +I(X_{2}^n;Y_{2}^n,\tilde{U}_{2}^n|Q)  \nonumber \\
  &\overset{(b)}{=} I({X}_{1}^n;{Y}_{1}^n|\tilde{U}_{1}^n,Q) +I(X_{2}^n;Y_{2}^n|\tilde{U}_{2}^n,Q) \nonumber \\
  &= \frac{1}{2}\Big\{ I({X}_{1}^n;{Y}_{1}^n|U_{1}^n)+I(X_{1}^n;Y_{1}^n)  \nonumber \\
  &\hspace{5mm} +I(X_{2}^n;Y_{2}^n|U_{2}^n) +I(X_{2}^n;Y_{2}^n) \Big\}
\end{align}
where $(a)$ follows from the independence between $X_i^n$ and $Q$, and $(b)$ is from the independence between $X_i^n$ and $\tilde{U}_{i}^n$. 
Similarly, we have
\begin{align} 
  n(R_1+R_2&-2\epsilon_n)   \nonumber \\
  &\le I({X}_{1}^n;{Y}_{1}^n,\tilde{S}_{1}^n) +I(X_{2}^n;Y_{2}^n,\tilde{U}_{2}^n)  \nonumber \\
  &\le \frac{1}{2}\Big\{I({X}_{1}^n;{Y}_{1}^n,S_{1}^n)+I(X_{1}^n;Y_{1}^n)  \nonumber \\
  &\hspace{5mm} +I(X_{2}^n;Y_{2}^n|U_{2}^n) +I(X_{2}^n;Y_{2}^n) \Big\} \label{eq:ob-7d} \\ 
  n(R_1+R_2&-2\epsilon_n) \nonumber \\
  &\le I({X}_{1}^n;{Y}_{1}^n,\tilde{U}_{1}^n) +I(X_{2}^n;Y_{2}^n,\tilde{S}_{2}^n)  \nonumber \\
  &\le \frac{1}{2}\Big\{I({X}_{1}^n;{Y}_{1}^n|U_{1}^n)+I(X_{1}^n;Y_{1}^n) \nonumber \\
  &\hspace{5mm} +I(X_{2}^n;Y_{2}^n,S_{2}^n) +I(X_{2}^n;Y_{2}^n) \Big\}. \label{eq:ob-7e}  
\end{align}

\addtocounter{equation}{1}%
\setcounter{storeeqcounter}{\value{equation}}

\begin{figure*}
  \normalsize
  \setcounter{tempeqcounter}{\value{equation}} 
  
  \begin{IEEEeqnarray}{ll}
\setcounter{equation}{\value{storeeqcounter}} 
  R_1+R_2 \;&\le \frac{1}{2}\Big\{h(Y_{1G})-h(Y_{1G}|X_{1G}) +h(U_{1G})-h(W_1-Z_1)  +h(Y_{1G}|U_{1G}) -h(h_{21}Y_{1G}+ \tilde{V}_{W_2}|U_{1G}) \nonumber \\ 
  &\hspace{9mm}  +h(Y_{2G})-h(Y_{2G}|X_{2G}) +h(U_{2G}) -h(W_2-Z_2) +h(Y_{2G}|U_{2G}) -h(h_{12}Y_{2G}+ \tilde{V}_{W_1}|U_{2G})\Big\} \label{eq:B-9} \\
     &{\ = \frac{1}{2}\Bigg\{\log \left(1+\frac{P_1}{|h_{12}|^2P_2+1}\right)   + \log \left(\frac{|h_{12}|^2P_2+\sigma_{W_1}^2}{\sigma_{W_1}^2+1-2\mathfrak{R}\{\rho_{W_1}\sigma_{W_1}\}}\right)} \nonumber \\ 
    &\hspace{9mm} + \log \left(\frac{P_1+|h_{12}|^2P_2+1-\frac{||h_{12}|^2P_2+\rho_{W_1}\sigma_{W_1}|^2}{|h_{12}|^2P_2+\sigma_{W_1}^2}}{|h_{21}|^2P_1+\sigma_{V_{W_2}}^2-\frac{|h_{21}\rho_{W_1}\sigma_{W_1}-h_{21}\sigma_{W_1}^2|^2}{|h_{12}|^2P_2+\sigma_{W_1}^2}}\right) \nonumber \\
    &\hspace{9mm} +\log \left(1+\frac{P_2}{|h_{{21}}|^2P_1+1}\right)   + \log \left(\frac{|h_{{21}}|^2P_1+\sigma_{W_{2}}^2}{\sigma_{W_{2}}^2+1-2\mathfrak{R}\{\rho_{W_{2}}\sigma_{W_{2}}\}}\right) \nonumber \\ 
    &\hspace{9mm} + \log \left(\frac{P_2+|h_{{21}}|^2P_1+1-\frac{||h_{{21}}|^2P_1+\rho_{W_{2}}\sigma_{W_{2}}|^2}{|h_{{21}}|^2P_1+\sigma_{W_{2}}^2}}{|h_{12}|^2P_2+\sigma_{V_{W_1}}^2-\frac{|h_{12}\rho_{W_{2}}\sigma_{W_{2}}-h_{12}\sigma_{W_{2}}^2|^2}{|h_{{21}}|^2P_1+\sigma_{W_{2}}^2}}\right) \vast\} \label{eq:B-10}
  \end{IEEEeqnarray}
  
  \setcounter{equation}{\value{tempeqcounter}} 
  \hrulefill
  \vspace*{6pt}
\end{figure*}

\begin{rem}  \normalfont
Although the time sharing parameter $Q$ also appears in several outer bounds on the capacity region (e.g., \cite{Tel07}), it was with respect to channel inputs (codes). The classical time sharing argument has been applied over different codes or rate pairs \cite{EK12}. 
\end{rem}

Fig. \ref{fig-b} illustrates the above genie-aided enhanced channel induced by \eqref{eq:ob-7d}.
} 

Even if one may find other useful combinations to get tighter sum-rate upper bounds, we restrict our attention to the above multi-letter expressions in this work. While we will derive a single-letter expression of (\ref{eq:ob-7c}) in Theorem \ref{thm-1}, two different single-letter expressions for both (\ref{eq:ob-7d}) and (\ref{eq:ob-7e}) will be given in  Theorems \ref{thm-5} and \ref{thm-6}.

\subsection{New Upper Bounds}
\label{sec:NB-B}

We begin with the multi-letter expression in (\ref{eq:ob-7c}) to derive the following result to upper-bound the sum rate $R_1+R_2$. 


\begin{thm} \label{thm-1} \normalfont
The capacity region of the two-user GIC  in the weak interference regime is contained in \eqref{eq:B-9}, 
shown on the top of page \pageref{eq:B-9},
\addtocounter{equation}{1}
for all $(W_1, W_2)$ satisfying 
\begin{align} 
  \sigma_{V_{W_1}}^{2} &\ge |h_{12}|^2\sigma^2_{Z_2-W_2} \label{eq:B-8b} \\
  \sigma_{V_{W_2}}^{2} &\ge |h_{21}|^2\sigma^2_{Z_1-W_1} \label{eq:B-8a} 
\end{align}
where $W_i, i=1,2,$ is given in (\ref{eq:A-4b}) and 
\begin{align} 
   \tilde{V}_{W_1}&=\sqrt{1-|h_{12}|^2\sigma_{V_{W_1}}^{-2}\sigma_{Z_2- W_2}^2}\;V_{W_1} \label{eq:B-8c}\\ 
   \tilde{V}_{W_2}&=\sqrt{1-|h_{21}|^2\sigma_{V_{W_2}}^{-2}\sigma_{Z_1- W_1}^2}\;V_{W_2} .\label{eq:B-8d}
\end{align}
\end{thm}

\begin{IEEEproof}
We first bound  $R_1$ by
\begin{align} \label{eq:B-21}
  n(R_1-\epsilon_n)&\le I(X^n_1; Y^n_1) \nonumber \\
 &\le nh(Y_{1G})-h(Y^n_1|X^n_1)  
\end{align}
and, similarly, $R_2$ by
\begin{align} \label{eq:B-22}
  n(R_2-\epsilon_n)&\le nh(Y_{2G})-h(Y^n_2|X^n_2). 
\end{align}
Using (\ref{eq:BT-4}), (\ref{eq:BT-4b}), (\ref{eq:B-21}), and (\ref{eq:B-22}), we can bound $R_1+R_2$ as follows:
\begin{align} 
  n&(2R_1+2R_2-4\epsilon_n)\nonumber \\ 
  &\le  nh(Y_{1G})-h(Y^n_1|X^n_1) +h(U^n_1)  - h(h_{12}X^n_2+V^n_{W_1})   \nonumber \\ &\ \ \ +h(Y^n_1|U^n_1)-nh(W_1-Z_1)  \nonumber \\ &\ \ \  +nh(Y_{2G})-h(Y^n_2|X^n_2) +h(U^n_2) - h(h_{21}X^n_1+V^n_{W_2})  \nonumber \\ &\ \ \  +h(Y^n_2|U^n_2)-nh(W_2-Z_2)  \nonumber \\ 
  &\overset{(a)}{\le} nh(Y_{1G})-nh(Y_{1G}|X_{1G}) +nh(U_{1G}) -nh(W_1-Z_1) \nonumber \\ &\ \ \   +h(Y^n_2|U^n_2)- h(h_{12}X^n_2+V^n_{W_1}) \label{eq:B-9c} \\ &\ \ \   +nh(Y_{2G}) -nh(Y_{2G}|X_{2G}) +nh(U_{2G}) -nh(W_2-Z_2) \nonumber \\ &\ \ \  +h(Y^n_1|U^n_1) - h(h_{21}X^n_1+V^n_{W_2})  \label{eq:B-9d}  \\ 
  &\overset{(b)}{\le} nh(Y_{1G})-nh(Y_{1G}|X_{1G}) +nh(U_{1G}) -nh(W_1-Z_1)  \nonumber \\ &\ \ \  +nh(Y_{2G}|U_{2G}) -nh(h_{12}Y_{2G}+ \tilde{V}_{W_1}|U_{2G}) \nonumber \\ &\ \ \  +nh(Y_{2G})-nh(Y_{2G}|X_{2G})  +nh(U_{2G}) -nh(W_2-Z_2) \nonumber \\ &\ \ \  +nh(Y_{1G}|U_{1G})  -nh(h_{21}Y_{1G}+ \tilde{V}_{W_2}|U_{1G})
\end{align}
where $(a)$ follows from applying Corollary \ref{lem-2} in Appendix \ref{sec:app-1} to 
\begin{align} 
  -h(Y^n_1|X^n_1) &+h(U^n_1)=\nonumber \\ 
  &-h(h_{12}X^n_2+Z^n_1) +h(h_{12}X^n_2+W^n_1) \nonumber \\
  -h(Y^n_2|X^n_2) &+h(U^n_2)=\nonumber \\ 
  &-h(h_{21}X^n_1+Z^n_2) +h(h_{21}X^n_1+W^n_2) \nonumber
\end{align} 
respectively, since $\sigma_{W_i}^2\le \sigma_{Z_i}^2=1$ for $i=1,2$, and $(b)$ immediately follows from applying Lemma \ref{lem-8} to 
\begin{align} 
  h(Y^n_2|U^n_2)- h&(h_{12}X^n_2+V^n_{W_1}) \nonumber \\
  =& \; h(X_2^n+h_{21}X^n_1+Z^n_2|h_{21}X^n_1+W^n_2) \nonumber \\
  &- h(h_{12}X^n_2+V^n_{W_1}) \nonumber \\
  =& \; h(X_2^n+h_{21}X^n_1+Z^n_2|h_{21}X^n_1+W^n_2)\nonumber \\
  &- h(X^n_2+h_{12}^{-1}V^n_{W_1}) -n\log|h_{12}| \nonumber \\
  h(Y^n_1|U^n_1)- h&(h_{21}X^n_1+V^n_{W_2}) \nonumber \\
  =&  \; h(X_1^n+h_{12}X^n_2+Z^n_1|h_{12}X^n_2+W^n_1) \nonumber \\
  &- h(h_{21}X^n_1+V^n_{W_2}) \nonumber \\
  =& \; h(X_1^n+h_{12}X^n_2+Z^n_1|h_{12}X^n_2+W^n_1) \nonumber \\
  &- h(X^n_1+h_{21}^{-1}V^n_{W_2}) -n\log|h_{21}| \nonumber
\end{align}
in (\ref{eq:B-9c}) and (\ref{eq:B-9d}) under the conditions in (\ref{eq:B-8b}) and (\ref{eq:B-8a}), respectively.
Then we obtain (\ref{eq:B-9}), yielding (\ref{eq:B-10}).
\end{IEEEproof}

Another upper bound on the sum rate $R_1+R_2$ {in the following theorem} will be found by using either (\ref{eq:ob-7d}) or (\ref{eq:ob-7e}), which is a \emph{hybrid} form of the Etkin-type genie-aided approach and the change-of-interference approach. 
Let the set of all parameters involved in the genie variables $S_i$ and the change-of-interference variables $U_i$ denoted as
\begin{align} \label{eq:NB-16}
  \boldsymbol{\kappa} \triangleq (\sigma_{N_1}, \sigma_{N_2}, \sigma_{W_1}, \sigma_{W_2}, \rho_{N_1}, \rho_{N_2}, \rho_{W_1}, \rho_{W_2})
\end{align}
where $0\le \sigma_{N_1}, \sigma_{N_2}, \sigma_{W_1}, \sigma_{W_2} \le 1$ and {$0\le|\rho_{N_1}|, |\rho_{N_2}|, |\rho_{W_1}|, |\rho_{W_2}| \le 1$}.
For the second upper bound, we need an intermediate step which is different from (\ref{eq:BT-1}) in the standard genie-aided approach. Starting from (\ref{eq:BT-1c}), we have
\begin{align} \label{eq:GO-1}
  n(R_1-\epsilon_n) &\le h(S^n_1)-nh(N_1) +h(Y^n_1|S^n_1) \nonumber \\
  &\hspace{5mm} -h(Y^n_1|X^n_1,S^n_1) \nonumber \\
  &= h(S^n_1)- nh(N_1) +h(Y^n_1|S^n_1) \nonumber \\
  &\hspace{5mm} -h(h_{12}X^n_2+V^n_{N_1})  
 \end{align}
 where $V_{N_1}^n$ is i.i.d. $\mathcal{N}(0,\sigma_{V_{N_1}}^2)$ with $\sigma_{V_{N_1}}^2=\sigma_{Z_1|N_1}^2$.
 Likewise, we have
\begin{align} \label{eq:GO-2}
  n(R_2-\epsilon_n) &\le h(S^n_2)- nh(N_2) +h(Y^n_2|S^n_2)  \nonumber \\
  &\hspace{5mm} -h(h_{21}X^n_1+V^n_{N_2}).
\end{align}
Then, we have the following upper bound on $R_1+R_2$.

\begin{thm} \label{thm-5} \normalfont
For all parameters $\boldsymbol{\kappa}$, defined in (\ref{eq:NB-16}), such that 
\begin{align}
  \boldsymbol{\kappa} \in \Big\{\boldsymbol{\kappa}: &\;\sigma_{V_{W_2}}^2\ge \min(\sigma_{N_1}^2, \sigma_{W_2}^2),\;\sigma_{V_{N_1}}^2 \ge\sigma_{Z_1-h_{21}^{-1}N_1}^2,  \nonumber \\
  &\ |h_{12}|^2\sigma_{Z_2-W_2}^2 \le 1\Big\}
\end{align} 
any rate pair $(R_1, R_2)$ in the capacity region of the two-user complex GIC  in the weak interference regime satisfies
\begin{align}
  &R_1+R_2  \nonumber \\
  &\ \le \frac{1}{2}\Big\{h(Y_{1G}|S_{1G}) -h(Y_{1G}+\tilde{V}_{N_1}|S_{1G}) \nonumber \\ 
  &\hspace{5mm} +h(Y_{2G}|U_{2G}) -h(h_{12}Y_{2G}+\tilde{Z}_1|U_{2G})+{\small\textsf{R}}_0\Big\} \label{eq:NB-9} \\  
  &\ = {\frac{1}{2}\vast\{ \log \left(\frac{P_1+|h_{12}|^2P_2+1-\frac{|h_{21}^*P_1+\rho_{N_1}\sigma_{N_1}|^2}{|h_{21}|^2P_1+\sigma_{N_1}^2}}{|h_{12}|^2P_2+\sigma_{V_{N_1}}^2-\frac{|\rho_{N_1}\sigma_{N_1}-h_{21}^{-1}\sigma_{N_1}^2|^2}{|h_{21}|^2P_1+\sigma_{N_1}^2}}\right)} \nonumber \\ 
&\hspace{4mm} +\log \left(\frac{P_2+|h_{21}|^2P_1+1-\frac{||h_{21}|^2P_1+\rho_{W_2}\sigma_{W_2}|^2}{|h_{21}|^2P_1+\sigma_{W_2}^2}}{|h_{12}|^2P_2+1-\frac{|h_{12}\rho_{W_2}\sigma_{W_2}-h_{12}\sigma_{W_2}^2|^2}{|h_{21}|^2P_1+\sigma_{W_2}^2}}\right)   +{\small\textsf{R}}_0 \vast\}  \label{eq:NB-9b}
\end{align}
where 
\begin{align}  
  \tilde{V}_{N_1}&=\sqrt{1-\sigma_{V_{N_1}^{}}^{-2}\sigma_{Z_1-h_{21}^{-1}N_1}^2}\;{V}_{N_1}\nonumber \\ 
   \tilde{Z}_1&=\sqrt{1-|h_{12}|^2\sigma_{Z_2-W_2}^2}\;Z_1 \nonumber
\end{align}  
and ${\small\textsf{R}}_0$ is given by
\begin{align}  \label{eq:NB-15}
  {\small\textsf{R}}_0= \;& h(S_{1G}) -h(h_{21}X_{1G}+V_{W_2}) +h(U_{2G}) -h(Y_{2G}|X_{2G}) \nonumber \\ 
  & \ \ \ +h(Y_{1G}) -h(N_1) +h(Y_{2G}) -h(Z_2-W_2) \nonumber \\
  =& \log \left(\frac{|h_{21}|^2P_1+\sigma_{N_1}^2}{|h_{21}|^2P_1+\sigma_{V_{W_2}}^2}\right) +\log \left(\frac{|h_{21}|^2P_1+\sigma_{W_2}^2}{|h_{21}|^2P_1+1}\right)  \nonumber \\ &\ \ \  +  \log \left(\frac{P_1+|h_{12}|^2P_2+1}{\sigma_{N_1}^2}\right)  \nonumber \\ &\ \ \ +\log \left(\frac{P_2+|h_{21}|^2P_1+1}{1+\sigma_{W_2}^2-2\mathfrak{R}\{\rho_{W_2}\sigma_{W_2}\}}\right).
\end{align}
Interchanging the user indices, we obtain another bound.
\end{thm}

\begin{IEEEproof}
Using (\ref{eq:BT-4b}),  (\ref{eq:B-21}), (\ref{eq:B-22}), and (\ref{eq:GO-1}), we can upper-bound the multi-letter expression in (\ref{eq:ob-7d}) as follows.
\begin{subequations}\label{eq:NB-11}
\begin{align} 
  n(2&R_1+2R_2-4\epsilon_n)  \nonumber \\ 
  &\le h(S^n_1)- nh(N_1) +h(Y^n_1|S^n_1)-h(h_{12}X^n_2+V^n_{N_1})  \nonumber \\ 
  &\ \ \ +nh(Y_{1G})-h(Y^n_1|X^n_1) \nonumber \\ 
  & \ \ \ +h(Y^n_2|U^n_2) - h(h_{21}X^n_1+V^n_{W_2}) +h(U^n_2)   \nonumber \\ 
  & \ \ \ -nh(Z_2-W_2)+nh(Y_{2G})-h(Y^n_2|X^n_2) \nonumber \\ 
  &= h(S^n_1) -h(h_{21}X^n_1+V^n_{W_2}) +h(U^n_2) -h(Y^n_2|X^n_2) \label{eq:NB-11a} \\ 
  & \ \ \ +h(Y^n_1|S^n_1) -h(h_{12}X^n_2+V^n_{N_1}) +h(Y^n_2|U^n_2) \nonumber \\ 
  & \ \ \ \ \ -h(Y^n_1|X^n_1)  \label{eq:NB-11b} \\
  & \ \ \ +nh(Y_{1G}) -nh(N_1) +nh(Y_{2G}) -nh(Z_2-W_2).  \label{eq:NB-11c}
\end{align}
\end{subequations}
We first bound (\ref{eq:NB-11a}) in the following two ways. Using the worst additive noise lemma, we can bound the first two terms in (\ref{eq:NB-11a}) as
\begin{align}  \label{eq:NB-3}
   h(S^n_1) -h&(h_{21}X^n_1+V^n_{W_2})  \nonumber \\ 
  & = h(h_{21}X_1^n+N_1^n) -h(h_{21}X^n_1+V^n_{W_2}) \nonumber \\ 
   &\le nh(h_{21}X_{1G}+N_1) -nh(h_{21}X_{1G}+V_{W_2}) \nonumber \\ 
   &= nh(S_{1G}) -nh(h_{21}X_{1G}+V_{W_2})
\end{align}
for $\sigma_{N_1}^2 \le \sigma_{V_{W_2}}^2$. Applying the worst additive noise lemma to $$h(U^n_2) -h(Y^n_2|X^n_2)=h(h_{21}X^n_1+W^n_2)- h(h_{21}X_1^n+Z_2^n)$$ 
due to the assumption that $\sigma_{W_2}^2 \le 1$, we also have 
\begin{align} \label{eq:NB-12}
  h(S^n_1&) -h(h_{21}X^n_1+V^n_{W_2}) +h(U^n_2) -h(Y^n_2|X^n_2) \nonumber \\ 
 &\le nh(S_{1G}) -nh(h_{21}X_{1G}+V_{W_2}) +nh(U_{2G}) \nonumber \\ 
  & \ \ \ -nh(Y_{2G}|X_{2G}).
\end{align}
The same single-letter expression as (\ref{eq:NB-12}) with a different condition on $\boldsymbol{\kappa}$ can be obtained by applying the worst additive noise lemma to 
\begin{align}
  h(U^n_2&) -h(h_{21}X^n_1+V^n_{W_2}) \nonumber \\ 
 &= h(h_{21}X^n_1+W^n_2)-h(h_{21}X^n_1+V^n_{W_2}) \nonumber
\end{align}
for $\sigma_{W_2}^2 \le \sigma_{V_{W_2}}^2$ and also to $$h(S^n_1) -h(Y^n_2|X^n_2)= h(h_{21}X_1^n+N_1^n) - h(h_{21}X_1^n+Z_2^n)$$ since $\sigma_{N_1}^2 \le 1$.  Thus, (\ref{eq:NB-12}) holds for $\sigma_{V_{W_2}}^2\ge \min(\sigma_{N_1}^2, \sigma_{W_2}^2)$.

Next, (\ref{eq:NB-11b}) can be upper-bounded by applying Lemma \ref{lem-8} to 
\begin{align}
   h(Y^n_1|S^n_1&) -h(h_{12}X^n_2+V^n_{N_1})  \nonumber \\
   &=h(h_{12}X^n_2+X^n_1+Z^n_1|h_{21}X_1^n+N_1^n) \nonumber \\
   &\ \ \ -h(h_{12}X^n_2+V^n_{N_1})  \nonumber
\end{align} 
for $\sigma_{Z_1-h_{21}^{-1}N_1}^2 \le \sigma_{V_{N_1}}^2$ and also  to 
\begin{align}
  h(Y^n_2|U^n_2&) -h(Y^n_1|X^n_1) \nonumber \\
   &=h(X_2^n+h_{21}X^n_1+Z_2^n|h_{21}X^n_1+W^n_2) \nonumber \\
   &\ \ \ -h(X^n_2+h_{12}^{-1}Z^n_1) -n\log|h_{12}| \nonumber 
\end{align}  
for $|h_{12}|^2\sigma_{Z_2-W_2}^2 \le 1$, we have 
\begin{align} \label{eq:NB-13}
  h(Y^n_1|S^n_1&) -h(h_{12}X^n_2+V^n_{N_1}) +h(Y^n_2|U^n_2) -h(Y^n_1|X^n_1) \nonumber \\ 
 &\le nh(Y_{1G}|S_{1G}) -nh(Y_{1G}+\tilde{V}_{N_1}|S_{1G}) \nonumber \\
   &\ \ \ +nh(Y_{2G}|U_{2G}) -nh(h_{12}Y_{2G}+\tilde{Z}_1|U_{2G}).
\end{align}
Using (\ref{eq:NB-11c}), (\ref{eq:NB-12}), and (\ref{eq:NB-13}), we obtain (\ref{eq:NB-9}).  
\end{IEEEproof}

\begin{rem}  \normalfont
Although we employ the genie signals in (\ref{eq:A-9}) to derive the new upper bound in Theorem \ref{thm-5}, we do not rely on the same bounding technique. Specifically, the central steps in the standard genie-aided approach are (\ref{eq:BT-1}) and (\ref{eq:BT-1b}). Our hybrid approach only makes use of (\ref{eq:BT-1c}), implying that we do not apply the standard argument that Gaussian maximizes entropy  to $h(Y^n_1|S^n_1)$ leading to (\ref{eq:BT-1}). Instead, we adopt the step in (\ref{eq:NB-13}), which was possible due to the conditional worst additive noise lemma.
\end{rem}


Meanwhile, we can derive different single-letter expressions of (\ref{eq:ob-7d}) and (\ref{eq:ob-7e}) by upper-bounding (\ref{eq:NB-11b}) in an alternative way, as shown by the following result.

\begin{thm} \label{thm-6} \normalfont
For all parameters $\boldsymbol{\kappa}$, defined in (\ref{eq:NB-16}), such that 
\begin{align} \label{eq:NB-10a}
   \boldsymbol{\kappa} \in \Big\{\boldsymbol{\kappa}: &\; \sigma_{V_{W_2}}^2\ge \min(\sigma_{N_1}^2, \sigma_{W_2}^2),\; \sigma_{V_{N_1}}^2\ge |h_{12}|^2\sigma_{Z_2-W_2}^2, \nonumber \\
   &\;\sigma_{Z_1-h_{21}^{-1}N_1}^2 \le 1\Big \}
\end{align}
any rate pair $(R_1, R_2)$ in the capacity region of the two-user GIC  in the weak interference regime satisfies
\begin{align} 
  &R_1+R_2\nonumber \\
   &\;\le \frac{1}{2}\Big\{h(Y_{1G}|S_{1G}) -h(Y_{1G}+\tilde{Z}'_1|S_{1G}) \nonumber \\ &\hspace{3mm} +h(Y_{2G}|U_{2G}) -h(h_{12}Y_{2G}+\tilde{V}'_{N_1}|U_{2G})+{\small\textsf{R}}_0\Big\} \label{eq:NB-10} \\
  &\;= {\frac{1}{2}\vast\{ \log \left(\frac{P_1+|h_{12}|^2P_2+1-\frac{|h_{21}^*P_1+\rho_{N_1}\sigma_{N_1}|^2}{|h_{21}|^2P_1+\sigma_{N_1}^2}}{|h_{12}|^2P_2+1-\frac{|\rho_{N_1}\sigma_{N_1}-h_{21}^{-1}\sigma_{N_1}^2|^2}{|h_{21}|^2P_1+\sigma_{N_1}^2}}\right) }
\nonumber \\ 
  &\hspace{3mm}   +\log \left(\frac{P_2+|h_{21}|^2P_1+1-\frac{||h_{21}|^2P_1+\rho_{W_2}\sigma_{W_2}|^2}{|h_{21}|^2P_1+\sigma_{W_2}^2}}{|h_{12}|^2P_2+\sigma_{V_{N_1}}^2-\frac{|h_{12}\rho_{W_2}\sigma_{W_2}-h_{12}\sigma_{W_2}^2|^2}{|h_{21}|^2P_1+\sigma_{W_2}^2}}\right) +{\small\textsf{R}}_0 \vast\} \label{eq:NB-10b}
\end{align}
where 
\begin{align} 
  \tilde{Z}'_1&=\sqrt{1-\sigma_{Z_1-h_{21}^{-1}N_1}^2}\;Z_1\nonumber \\ 
  \tilde{V}'_{N_1}&=\sqrt{1-\sigma_{V_{N_1}^{}}^{-2}|h_{12}|^2\sigma_{Z_2-W_2}^2}\;{V}_{N_1}. \nonumber
\end{align} 
Interchanging the user indices, we obtain another bound.
\end{thm}

\begin{figure}
\hspace{-5mm}
  \includegraphics[scale=.51]{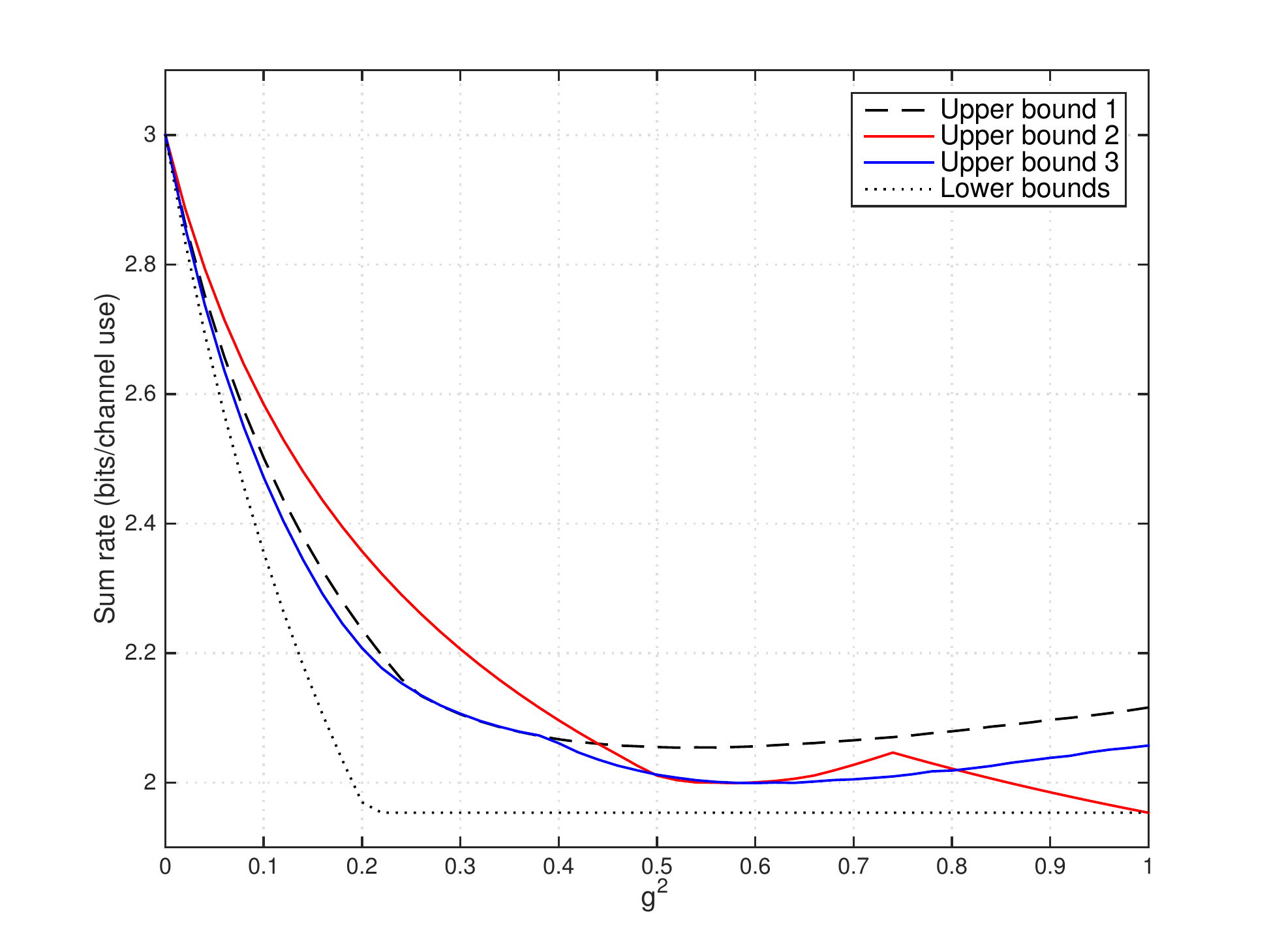}
  \caption{Bounds on the sum capacity of the symmetric Gaussian interference channel for $P=7$. Upper bounds 1, 2, 3 are calculated from Theorems \ref{thm-1}, \ref{thm-5}, \ref{thm-6}, respectively. The lower bound is given  by the maximum of the treating-interference-as-noise (TIN) and the TDM/frequency division multiplexing (FDM) scheme.}\label{fig-1}
\end{figure}

\begin{figure*}
\hspace{-7mm}
\center
  \includegraphics[scale=.75]{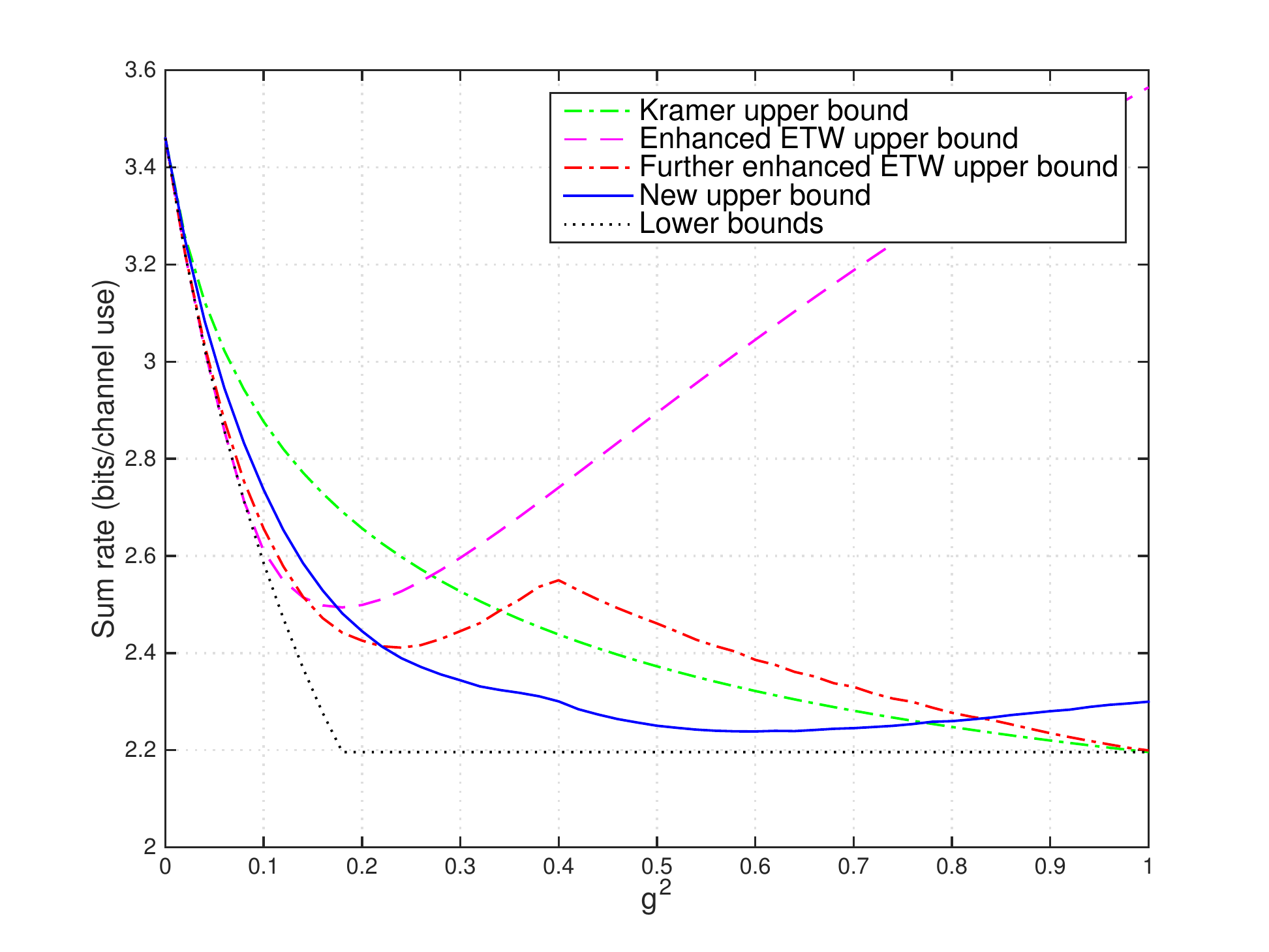} 
\caption{Bounds on the sum capacity of two-user symmetric Gaussian interference channel for $P=10$. {The Kramer upper bound is given by \cite[Thm. 2]{Kra04}. The new upper bound here is calculated by minimizing (\ref{eq:NB-10b}).} 
  }\label{fig-2}
\end{figure*}

\begin{IEEEproof}
Similar to (\ref{eq:NB-13}), applying Lemma \ref{lem-8} to 
\begin{align} 
  h&(Y^n_1|S^n_1) -h(Y^n_1|X^n_1) \nonumber \\
   &=h(h_{12}X^n_2+X_1^n+Z^n_1|h_{21}X_1^n+N_1^n)-h(h_{12}X^n_2+Z^n_1) \nonumber
\end{align}
for $\sigma_{Z_1-h_{21}^{-1}N_1}^2 \le 1$ and also to 
\begin{align} 
  &h(Y^n_2|U^n_2) -h(h_{12}X^n_2+V^n_{N_1}) \nonumber \\
   &\;=h(X_2^n+h_{21}X^n_1+Z_2^n|h_{21}X^n_1+W^n_2) -h(h_{12}X^n_2+V^n_{N_1}) \nonumber
\end{align}
for $|h_{12}|^2\sigma_{Z_2-W_2}^2 \le \sigma_{V_{N_1}}^2$, we get a different bound to (\ref{eq:NB-11b}) as follows: 
\begin{align} \label{eq:NB-14}
  h(Y^n_1&|S^n_1) -h(Y^n_1|X^n_1) +h(Y^n_2|U^n_2) -h(h_{12}X^n_2+V^n_{N_1}) \nonumber \\ 
 &\le nh(Y_{1G}|S_{1G}) -nh(Y_{1G}+\tilde{Z}'_1|S_{1G}) \nonumber \\
   &\ \ \ +nh(Y_{2G}|U_{2G}) -nh(h_{12}Y_{2G}+\tilde{V}'_{N_1}|U_{2G}).
\end{align}
Combining (\ref{eq:NB-11c}), (\ref{eq:NB-12}), and (\ref{eq:NB-14}), we have (\ref{eq:NB-10}).
\end{IEEEproof}

{The main difference between Theorems \ref{thm-5} and  \ref{thm-6} is that the latter relates a pair of differential entropies such that $h(Y^n_2|U^n_2) -h(Y^n_1|X^n_1,S^n_1)\le nh(Y_{2G}|U_{2G}) -nh(h_{12}Y_{2G}+\tilde{V}'_{N_1}|U_{2G})$, which will become $n\log|g|^{-2}$ in the next section. The $n\log|g|^{-2}$ term will be shown therein to be essential to characterize the high-SNR capacity of the symmetric GIC. 
}

In order to show how useful the new sum-rate upper bounds are, we provide numerical results for the symmetric real GIC in (\ref{eq:A-1b}). 
Fig. \ref{fig-1} depicts the sum-rate upper bounds in Theorems \ref{thm-1},  \ref{thm-5}, and  \ref{thm-6} (indicated by upper bounds 1, 2, and 3, respectively) for the two-user symmetric real GIC when $P=7$. The lower bound is given by the maximum of the treating-interference-as-noise and the time division bounds. We can see that upper bound 3 in (\ref{eq:NB-10}) coincides with bound 1 in (\ref{eq:B-9}) over a certain range of $g^2$, while in general it is slightly outperformed only for a very narrow range of $g^2$ by bound 2 in (\ref{eq:NB-9}), which was consistently observed in most cases of interest. {Accordingly, we will only consider upper bound 3 in Theorem \ref{thm-6} throughout the remaining sections of this paper. }

Fig. \ref{fig-2} compares our upper bound in Theorem \ref{thm-6} with other known upper bounds for $P=10$, {where the Kramer upper bound is taken from \cite[Thm. 2]{Kra04}, the enhanced ETW bound is from the minimum over the bounds in \cite{Mot09,Sha09,Ann09}, and the further enhanced ETW bound in \cite{Etk09} is also plotted. Our upper bound in Theorem \ref{thm-6} is shown to be tighter than the existing bounds at a certain range of $g^2$.}


\section{Simplified Capacity Bounds for the Symmetric Real Case} 
\label{sec:NB-A}

In this section, we consider the {symmetric real} GIC with constant channel coefficients in (\ref{eq:A-1b}) to  simplify the upper and lower bounds on the sum capacity. 
Notice that all the proposed bounds in the previous section should be numerically found by minimizing over the appropriate ranges of all relevant parameters. In order to avoid this numerical optimization, we will simplify the upper bounds for the symmetric real GIC. For this, note that the equalities in the worst additive noise lemma in \cite{Dig01} and its conditional version in Lemma \ref{lem-4} trivially hold true when $\sigma_Z^2=0$ as well as when $X^n=X_g^n$ (and $U^n=U_g^n$ for Lemma \ref{lem-4}). Therefore, those lemmas incur no loss in tightness of the resulting bounds if one can safely let $\sigma_Z^2=0$. 
The following result is a simplified upper bound on the sum rate $R$ based on Theorem \ref{thm-6}.

\begin{thm} \label{thm-7} \normalfont
The sum rate of the symmetric real GIC for $0<g^2\le 1$ is bounded by  
\begin{align} \label{eq:NB-20}
  R \le &R_\text{sym}^* + \min\bigg(\frac{1}{4}\log {\frac{g^2P+\sigma_{N_1}^2}{g^2P+1}} \nonumber \\&\ \ + \frac{1}{4}\log \frac{4g^4}{\sigma_{N_1}^2(4g^2-\sigma_{N_1}^2)}, \ \frac{1}{2}\log \frac{4g^2+1}{4|g|}\bigg)
\end{align}
where $R_\text{sym}^*$ is given in (\ref{eq:intro-1}), and 
\begin{align} \label{eq:NB-31}
  \sigma_{N_1}^2 = \left\{ \begin{array}{ll}
  4g^2({1-g^2}) &  \ \text{if } g^2\le 0.5\\
  1 &  \ \text{otherwise}.
  \end{array} \right.
\end{align}
\end{thm}

\begin{IEEEproof}
The upper bound in (\ref{eq:NB-10}) can be naturally simplified with the equalities in the conditions in (\ref{eq:NB-10a}) due to the use of the worst additive noise lemma and its conditional version.  In the symmetric case, $\boldsymbol{\kappa}$    in (\ref{eq:NB-16}) reduces to  $\boldsymbol{\kappa}_0=(\sigma_{N_1}, \sigma_{W_2}, \rho_{N_1}, \rho_{W_2})$. However, we cannot simultaneously satisfy all the constraints in  (\ref{eq:NB-10a}) and there might be a number of ways to simplify the symmetric-rate upper bound based on (\ref{eq:NB-10}). In the sequel, we will only consider two special cases. 
Given a fixed $\sigma_{N_1}$, we begin with 
\begin{enumerate}
\item[A1)] Let $\sigma_{Z_1-g^{-1}N_1}^2=1$ (i.e., $\tilde{Z}'_1=0$) to meet $h(Y_{1G}|S_{1G}) -h(Y_{1G}+\tilde{Z}'_1|S_{1G})=0$ in (\ref{eq:NB-10}). Since $\sigma_{Z_1-g^{-1}N_1}^2= 1+g^{-2}\sigma_{N_1}^2-2g^{-1}\sigma_{N_1}\rho_{N_1}$, we have 
\begin{align} \label{eq:NB-22}
   \rho_{N_1}=\frac{\sigma_{N_1}}{2g}.
\end{align}
\item[A2)] Let $\sigma_{V_{W_2}}^{2}=\sigma_{W_2}^2$ to meet $h(U_{2G}) -h(gX_{1G}+V_{W_2})=0$ in (\ref{eq:NB-15}). Since $\sigma_{V_{W_2}}^{2}=1-\frac{(1-\sigma_{W_2}\rho_{W_2})^2}{1+\sigma_{W_2}^2-2\sigma_{W_2}\rho_{W_2}}$, we have 
\begin{align} \label{eq:NB-27}
   \sigma_{W_2}^2 = \rho_{W_2}^2.
\end{align}
\item[A3)] Let $\sigma_{V_{N_1}}^{2}=g^2\sigma_{Z_2-W_2}^2$ (i.e., $\tilde{V}'_{N_1}=0$) to meet $h(Y_{2G}|U_{2G}) -h(gY_{2G}+\tilde{V}'_{N_1}|U_{2G})=-\frac{1}{2}\log |g|^2$ in (\ref{eq:NB-10}). 
Given (\ref{eq:NB-27}), we have 
\begin{align} \label{eq:NB-25}
   \rho_{W_2} 
   =g^{-1}\sqrt{g^2+\rho_{N_1}^2-1}
\end{align}
since $\sigma_{V_{N_1}}^{2}=1-\rho_{N_1}^2$ and $\sigma_{Z_2-W_2}^2={1+\sigma_{W_2}^2-2\sigma_{W_2}\rho_{W_2}}$.
\end{enumerate}
Plugging (\ref{eq:NB-22}), (\ref{eq:NB-27}), and (\ref{eq:NB-25}) into (\ref{eq:NB-10}), we get  
\begin{align} \label{eq:NB-29}
  R &\le \min_{\sigma_{N_1}^2} \frac{1}{2}\log (P+g^2P+1) -\frac{1}{4}\log |g|^2 \nonumber \\
  &\hspace{5mm} +\frac{1}{4}\log {\frac{g^2P+\sigma_{N_1}^2}{g^2P+1}} + \frac{1}{4}\log \frac{4g^4}{\sigma_{N_1}^2(4g^2-\sigma_{N_1}^2)}
\end{align}
subject to  
\begin{align} \label{eq:NB-30}
  {4g^2{(1-g^2)}\le\sigma_{N_1}^2\le\min(1,4g^2)}.
\end{align}
At high SNR, $\log {\frac{g^2P+\sigma_{N_1}^2}{g^2P+1}}\approx 0$. Finding $\sigma_{N_1}^2$ that minimizes (\ref{eq:NB-29}) can then be simplified by 
$$\min_{\sigma_{N_1}^2}  \log \frac{4g^4}{\sigma_{N_1}^2(4g^2-\sigma_{N_1}^2)}$$
for $\sigma_{N_1}^2$ satisfying (\ref{eq:NB-30}). With the solution given in (\ref{eq:NB-31}), we can further upper-bound (\ref{eq:NB-29}) as 
\begin{align} \label{eq:NB-29b}
  R &\le  \frac{1}{2}\log (|g|P+|g|^{-1}(P+1)) + \frac{1}{4}\log {\frac{g^2P+\sigma_{N_1}^2}{g^2P+1}} \nonumber \\
  &\hspace{5mm} + \frac{1}{4}\log \frac{4g^4}{\sigma_{N_1}^2(4g^2-\sigma_{N_1}^2)}.
\end{align}


We can satisfy the equalities in the constraints in Theorem \ref{thm-6} with a set of different combinations of positive and negative conditional entropies. Keeping (\ref{eq:NB-22}) in step A1), we next forces $h(U_{2G}) -h(Y_{2G}|X_{2G})=0$ instead of $h(U_{2G}) -h(gX_{1G}+V_{W_2})=0$ in step A2) as follows: 
\begin{enumerate}
\item[B1)] $h(U_{2G}) -h(Y_{2G}|X_{2G})=0$ in (\ref{eq:NB-15}) is simply done by 
\begin{align} \label{eq:NB-28}
   \sigma_{W_2}^2=1.
\end{align} 
\item[B2)] Let $\sigma_{V_{N_1}}^{2}=g^2\sigma_{Z_2-W_2}^2$ again, but in the following way unlike A3). 
Given (\ref{eq:NB-22}) and (\ref{eq:NB-28}), we can do so by 
\begin{align} \label{eq:NB-26}
   \rho_{W_2}=\frac{\sigma_{N_1}^2}{8g^4}-\frac{1}{2g^2}+1.
\end{align}
\item[B3)] Let $\sigma_{V_{W_2}}^{2}=\sigma_{N_1}^2$ to meet $h(S_{1G}) -h(gX_{1G}+V_{W_2})=0$ in (\ref{eq:NB-15}). Given (\ref{eq:NB-28}) and (\ref{eq:NB-26}), we have
\begin{align} \label{eq:NB-34}
  \sigma_{N_1}^2 = \frac{4g^2}{{4g^2+1}}.
\end{align}
\end{enumerate}
Plugging (\ref{eq:NB-22}), (\ref{eq:NB-28}), (\ref{eq:NB-26}), and (\ref{eq:NB-34}) into (\ref{eq:NB-10}) leads to another upper bound 
\begin{align} \label{eq:NB-32b}
  R \le \frac{1}{2}& \log \big(|g|P+|g|^{-1}(P+1)\big) +  \frac{1}{2}\log \frac{4g^2+1}{4|g|}.
\end{align}
Combining (\ref{eq:NB-29b}) and (\ref{eq:NB-32b}) completes the proof. 
\end{IEEEproof}


\begin{rem} \normalfont
The difference between the upper bounds in Theorems \ref{thm-6} and \ref{thm-7} may be even arbitrarily large  when interference is very weak and SNR is low, i.e., outside the moderate interference regime. However, the difference turns out to be negligible inside the regime of our interest. We skip plotting the difference for the sake of compactness of this work.
\end{rem}


We can further simplify the upper bound in Theorem \ref{thm-7} to get a more desirable closed-form bound. By letting $\sigma_{N_1}=1$ and seeing that $\frac{4g^2+1}{4|g|}=\frac{2g^2}{\sqrt{4g^2-1}}$ when $g^2 = 0.405\cdots$, we have the following corollary. 

\begin{cor} \label{cor-1} \normalfont
The sum rate of the symmetric real GIC is upper-bounded by  
\begin{align} \label{eq:NB-20b}
  \overline{R} = R_\text{sym}^* +\gamma(g)
\end{align}
where $\gamma(g)$ is given by
\begin{align} \label{eq:NB-24}
  \gamma(g) = \left\{ \begin{array}{ll}
  \frac{1}{2}\log \frac{4g^2+1}{4|g|} &  \ 0< g^2 \le 0.405\\
  \frac{1}{2}\log \frac{2g^2}{\sqrt{4g^2-1}} &  \ 0.405 < g^2\le 1 
  \end{array} \right. .
\end{align}
\end{cor}

It is easy to see that $\overline{R} \ge R_\text{sym}^*$ since $\gamma(g)\ge 0$ for all $g$.
In what follows, we investigate the rate gap between the HK lower bound and the simplified upper bound $\overline{R}$ on the capacity of the symmetric real GIC in the moderate interference regime. 
Furthermore, the capacity of the symmetric real GIC will be characterized to within 0.104 bits/Hz by $R_\text{sym}^*$. 

We first simplify the well-known special cases of the HK scheme in \cite{Etk08,Sas04}.  
For $P^{-1/3}<g^2<1$, it can be shown that if $b=0$ and $c=0.5$,\footnote{In this work, we replace the notations of \cite{Sas04} with $\alpha=a, \beta=b,\delta=c,$ and $\bar{a}=1-a$. Clearly, making $b$ and $c$ fixed does not increase the HK lower bound.} the second term of the $\min(\cdot,\cdot,\cdot)$ of the HK scheme with $|Q|=4$ in \cite[Eq. (32)]{Sas04} becomes inactive, where $Q$ is the time sharing parameter.  The resulting sum rate  (denoted by $R_\text{HK}$) then  reduces  to
\begin{align} \label{eq:NB-40}
  R_\text{HK} = &\max_{a\in [0,1]}\frac{1}{2} \log(1+aP) +\min \Bigg\{\frac{1}{4} \log \Big(1+\frac{\bar{a}P+g^2P}{1+aP}\Big) \nonumber \\
  &\hspace{5mm} + \frac{1}{4} \log\Big(1+\frac{P+g^2\bar{a}P}{1+g^2aP}\Big), \nonumber \\
  &\ \ \frac{1}{2} \log\Big(1+\frac{g^2\bar{a}P}{1+g^2aP}\Big) +\frac{1}{2} \log \Big(1+\frac{g^2P}{1+aP}\Big) \Bigg\} .
\end{align}
The above max-min problem is solved by 
\begin{align} \label{eq:NB-39}
  a^* = \frac{a_2+\sqrt{a_2^2-4(a_0g^2-a_1)(a_0-a_1^2)}}{2(a_0g^2-a_1)P}
\end{align}
where $a_0=(1+P+g^2P)^2, a_1=(1+g^2P)^2,$ and $a_2=2a_1^{3/2}-a_0(1+g^2)$. Then 
\begin{align} \label{eq:NB-36c}
  R_\text{HK} &= \frac{1}{2} \log(1+a^*P) +\frac{1}{4} \log \Big(1+\frac{\bar{a^*}P+g^2P}{1+a^*P}\Big)\nonumber \\
  &\hspace{5mm}  + \frac{1}{4} \log\Big(1+\frac{P+g^2\bar{a^*}P}{1+g^2a^*P}\Big)  \nonumber \\
  &= R_\text{sym}^*  +\frac{1}{4} \log \frac{1+{a^*}P}{g^{-2}+a^*P} .
\end{align}

\begin{figure*}
\hspace{-5mm} \center
  \includegraphics[scale=1.0]{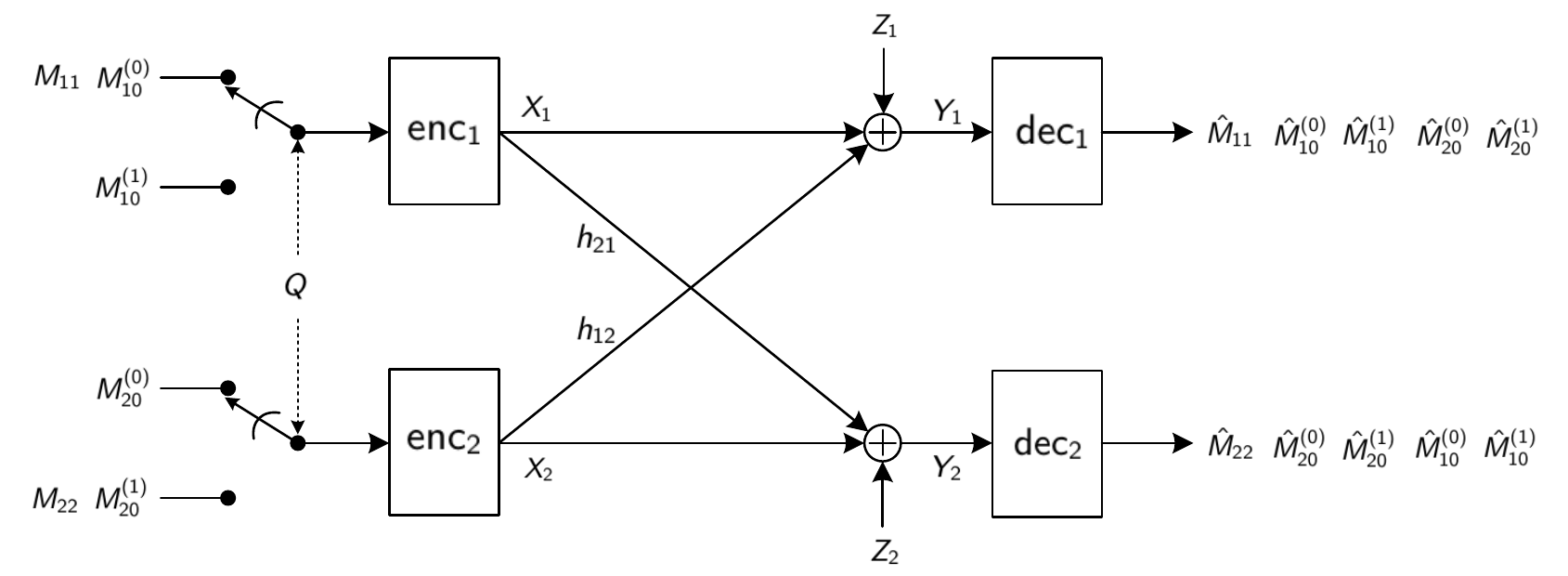}
  \caption{The special case in \eqref{eq:NB-36c} of the Han-Kobayashi achievable scheme. Sender 1 (sender 2)  transmits its private message $M_{11}$ ($M_{22}$) only when $Q=0$ ($Q=1$) with $\Pr(Q=0)=\frac{1}{2}$ ($\Pr(Q=1)=\frac{1}{2}$).}\label{fig-c}
\end{figure*}

The condition $P^{-1/3}<g^2\le1$ implies $\frac{2}{3}<\alpha\le 1$. At high SNR, the sum rate of the ``simplified HK scheme" in \cite[Eq. 4]{Etk08} (denoted by $R_\text{SHK}$) then reduces to 
\begin{align} \label{eq:NB-37}
  R_\text{SHK}  &= \frac{1}{2} \log(1+P+g^2P) +\frac{1}{2} \log (2+g^{-2}) -1 \nonumber \\
  &= R_\text{sym}^* +\frac{1}{2} \log \frac{2|g|+|g|^{-1}}{4}.
\end{align}

Even if the HK scheme in \cite{Han81} includes the TDM/FDM scheme as a special case, the above simplified HK schemes are generally not the case. Noticing that 
\begin{align} \label{eq:NB-39b}
  a^* = \frac{1+|g|+g^2}{1+g^2+g^4}|g|^3 +\Oc(P^{-1}) \ge |g|^3 +\Oc(P^{-1})
\end{align} 
we let $a=|g|^3$ and denote the resulting rate by  
\begin{align} \label{eq:NB-36a}
  \underline{R}_\text{HK} \triangleq R_\text{sym}^*  +\frac{1}{4} \log \frac{1+|g|^3P}{g^{-2}+|g|^3P}  .
\end{align}
For $P=23.239\cdots$ and $g^2=P^{-1/3}$, we can see $\underline{R}_\text{HK}= R_\text{TDM}$, where 
\begin{align} \label{eq:NB-37b}
    R_\text{TDM}= \frac{1}{2} \log(1+2P)
\end{align}
 is the achievable rate of the TDM scheme with power control. Since $R_\text{sym}^* > R_\text{TDM}$ for $0<g^2\le 1$ and $\frac{1}{4} \log \frac{1+|g|^3P}{g^{-2}+|g|^3P}\le 0$ in (\ref{eq:NB-36a}) is monotonically increasing with $P$, we get
$$\underline{R}_\text{HK}> R_\text{TDM}$$ for $P\ge 23.3$ and $P^{-1/3}<g^2<1$. Moreover, given $P\ge 23.3$ with $P^{-1/3}<g^2<1$, it is straightforward to see $a^*\ge |g|^3$. 
Then we also have $$R_\text{HK} \ge \underline{R}_\text{HK}$$ since (\ref{eq:NB-36c})  is monotonically increasing with $a^*$.
As a result, the simplified lower bound on the sum capacity (denoted by $\underline{R}$) to be considered in this work is given by 
\begin{align} \label{eq:NB-44}
\underline{R} \triangleq \left\{ \begin{array}{ll}
  R_\text{TDM}  &  \ P< 23.3 \\
  \max(\underline{R}_\text{HK},R_\text{SHK})  &   \  \text{otherwise}.
  \end{array} \right.
\end{align}

Define the rate gap $\Delta $ between the simplified new upper bound and the simplified lower bound as 
$$\Delta \triangleq \overline{R} - \underline{R}$$ where $\overline{R}$ and $\underline{R}$ are given by (\ref{eq:NB-20b}) and (\ref{eq:NB-44}), respectively. Eventually, we have the following result.

\begin{thm} \label{thm-8} \normalfont
In the moderate interference regime, where $P^{-1/3}<g^2\le1$, the rate gap $\Delta$ between the lower and the new upper bound on the capacity of the symmetric real GIC is upper-bounded as follows: For $P< 23.3$  
\begin{align} \label{eq:NB-41}
  \Delta \le \frac{1}{2}& \log \frac{|g|P+|g|^{-1}(P+1)}{1+2P} +\gamma(g) 
\end{align}
where $\gamma(g)$ is given in (\ref{eq:NB-24}).
Otherwise 
\begin{align} \label{eq:NB-42}
  \Delta \le &\max \bigg( \frac{1}{2} \log \frac{4}{2|g|+|g|^{-1}}, \ 
  \frac{1}{2}\log \frac{g^{-2}+|g|^3P}{1+|g|^3P} \bigg) \nonumber \\
  &\ \  +\gamma(g)  .
\end{align}
\end{thm}


\begin{rem} \label{rem-5} \normalfont
Let $M_{ii}$ denote the private message of sender $i=1,2$ and $M_{i0}$ denote the common message. 
It is important to notice that ${R}_\text{HK}$ in \eqref{eq:NB-36c} comes from 
\begin{align}
 I(Y_{1}; \mathcal{U}_{1}|&\mathcal{W}_{1},\mathcal{W}_{2},Q) + I(Y_{2}; \mathcal{U}_{2}|\mathcal{W}_{1},\mathcal{W}_{2},Q) \nonumber \\
  &+ I(Y_{1}; \mathcal{W}_{1},\mathcal{W}_{2}|Q) \nonumber
\end{align} 
\begin{figure*}
\hspace{-7mm}
\center
  \includegraphics[scale=.75]{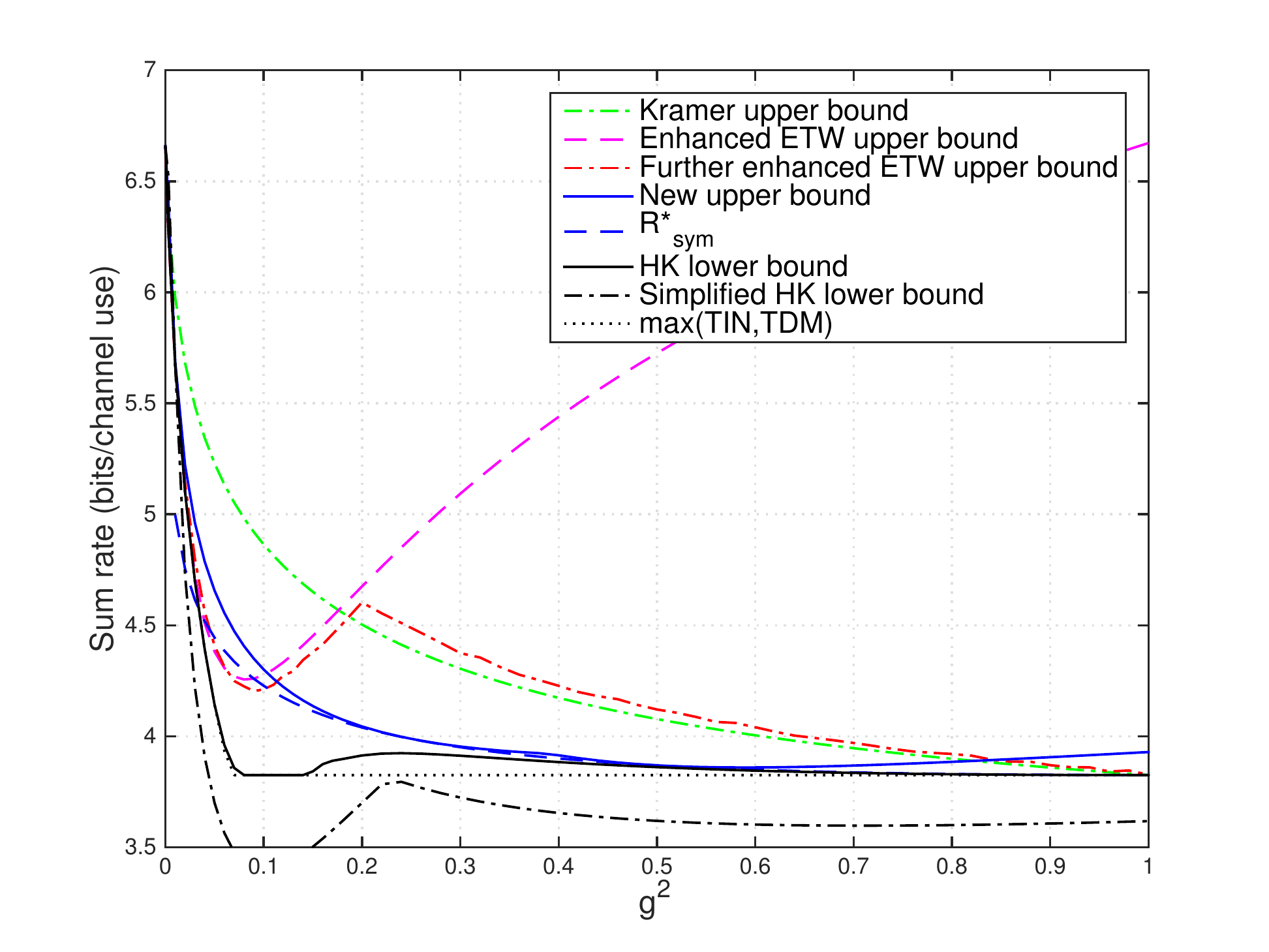}
  \caption{Bounds on the sum capacity of the symmetric GIC at $P=100$ (i.e., SNR $=20$ dB). {The new upper bound is calculated from the closed-form expression in (\ref{eq:NB-20}). $R_\text{sym}^*$ is given by (\ref{eq:intro-1}). The HK lower bound is given by (32) in \cite{Sas04}. The simplified HK bound is given by (4) in \cite{Etk08}. $\max$(TIN, TDM) is given by the maximum of the TIN and the TDM/FDM scheme.} }\label{fig-3a}
\end{figure*}
in \cite[Eq. 26]{Sas04},\footnote{By symmetry, we have the equivalent mutual information by interchanging the user indices.} where the random variables $\mathcal{U}_i, \mathcal{W}_i$ represent (i.e., encode) the private and common messages $M_{ii}, M_{i0}$ of user $i=1,2$, respectively, with the specific setting of the time sharing parameter $Q$  in Table I  in \cite{Sas04} while $Q=0,1$ are only active. Fig. \ref{fig-c} illustrates this special case of the HK scheme, where the superscript $^{(j)}$ indicates $Q=j$  with $|Q|=2$. The particular HK scheme that achieves ${R}_\text{HK}$ uses a single private message at either transmitter and two common messages. To be specific, with probability $1/2$ only transmitter 1 transmits its own private message represented by the Gaussian codeword $\mathcal{U}_{1}\sim \mathcal{N}(0,|g|^3P)$ and common messages represented by $\mathcal{W}_{1}\sim \mathcal{N}(0,(1-|g|^3)P)$ and $\mathcal{W}_{2}\sim \mathcal{N}(0,P)$ form multiple access channel (MAC) at receiver 1, while with the remaining probability only transmitter 2 transmits the private message represented by $\mathcal{U}_{2}\sim \mathcal{N}(0,|g|^3P)$ and common messages represented by $\mathcal{W}_{1}\sim \mathcal{N}(0,P)$ and $\mathcal{W}_{2}\sim \mathcal{N}(0,(1-|g|^3)P)$ form MAC at receiver 1. We do not require the two common messages to form MAC at receiver 2. Each receiver simultaneously decodes the common messages first while treating the single private message as noise.  
Even if the simplified HK scheme in \cite{Etk08} without time sharing is at most 1 bit away from the sum capacity, as earlier mentioned in \cite{Sas04}, the use of time sharing is very relevant in the high SNR and moderate interference regime to reduce the rate gap. 

\end{rem}

Let $\Cc_\text{sym}(P,g)$ denote the sum capacity of the symmetric real GIC.  The above result leads to the following important corollary. 

\begin{cor} \label{cor-3} \normalfont
For $g^2\in(\max(0.086, P^{-1/3}),1]$, the capacity of the symmetric real GIC is characterized by  
\begin{align} \label{eq:NB-45}
  \big|\;\Cc_\text{sym}(P,g) - R_\text{sym}^*\big| \le \frac{1}{2}\log\frac{2}{\sqrt{3}} < 0.104 .
\end{align}
\end{cor}

\begin{IEEEproof}
To see (\ref{eq:NB-45}), it suffices to show $\overline{R}-R_\text{sym}^*\le \frac{1}{2}\log\frac{2}{\sqrt{3}}$ and $\underline{R}-R_\text{sym}^*\ge  -\frac{1}{2}\log\frac{2}{\sqrt{3}}$. Noticing that the two functions of $g$ in (\ref{eq:NB-24}) are convex on their own intervals of $g$, we only need to calculate the values of $\overline{R}$ at $g^2=\max(0.086, P^{-1/3}), 0.405, 1$. Then we can see $\overline{R}-R_\text{sym}^*\le \frac{1}{2}\log\frac{2}{\sqrt{3}}$, where the equality holds when $g^2=1$, and $\frac{1}{2}\log\frac{2}{\sqrt{3}}=0.1037\cdots<0.104$. 

In order to lower-bound $\underline{R}-R_\text{sym}^*$, we first consider $R_\text{TDM}$ that becomes $\underline{R}$ for $P<23.3$ as in (\ref{eq:NB-44}). We can see that $R_\text{TDM}-R_\text{sym}^*$ has the largest value of $0.1023\cdots$ at $P=23.3$. Then, it suffices to lower-bound $\underline{R}_\text{HK}-R_\text{sym}^*$. We notice that 
\begin{align} 
  \argmin_{P^{-1/3}\le g^2<1}\underline{R}_\text{HK}-R_\text{sym}^*&=\argmin_{P^{-1/3}\le g^2<1} \frac{1}{4} \log \frac{1+|g|^3P}{g^{-2}+|g|^3P} \nonumber \\
  &=P^{-1/3} 
\end{align}   
since the second term in $\underline{R}_\text{HK}$ in (\ref{eq:NB-36a}) is monotonically increasing with $g$. Then the solution that minimizes the above problem is $g^2=0.25$. 
Thus $\underline{R}_\text{HK}-R_\text{sym}^*\ge -\frac{1}{2}\log\frac{\sqrt{3}}{2}$, where the equality holds when $P=g^{-6}=64$. 
\end{IEEEproof}

\begin{figure*}
\hspace{-7mm}
\center
  \includegraphics[scale=.75]{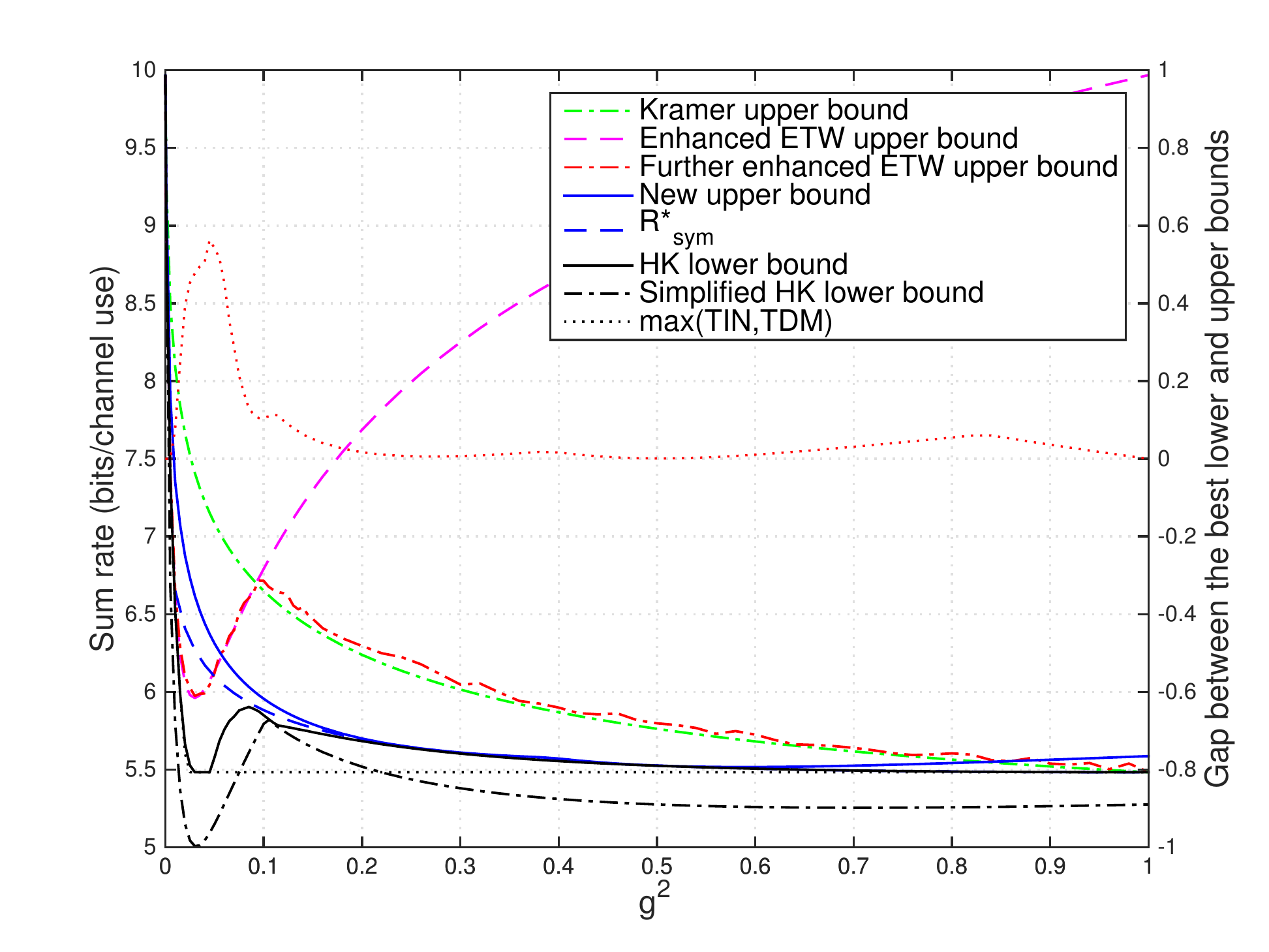}
  \caption{Bounds on the sum capacity of the symmetric GIC at $P=1000$ (i.e., SNR $=30$ dB). The red dotted curve indicates the smallest gap (in bits/channel use) between lower and upper bounds in the second y-axis on the right hand side. The remaining curves are calculated with the same methods as Fig. \ref{fig-3a}.}\label{fig-3b}
\end{figure*}

As a matter of fact, we have restricted our attention to the condition $\max(0.086, P^{-1/3})<g^2$ in the moderate interference regime to prevent $\underline{R}_\text{HK}$ (rather than $\overline{R}$) from being arbitrarily loose when interference is very weak and SNR is high (i.e., $P>10^3$). We will see in Corollary \ref{cor-2} that $R_\text{SHK}>\underline{R}_\text{HK}$ for $g^2\le 0.086$ at high SNR. 

\begin{cor} \label{cor-4} \normalfont
The simplified lower bound $\underline{R}$ in (\ref{eq:NB-44}) is optimal to within $0.125$ bit in the moderate interference regime. 
\end{cor}

\begin{IEEEproof}
The proof immediately follows from Corollaries \ref{cor-1} and \ref{cor-3}. Namely, 
$\underline{R}$ is at most $0.104$ bit below from $R_\text{sym}^*$ and $\overline{R}$ is at most $\gamma(g)$ bit above from $R_\text{sym}^*$. Since $\gamma(g)= 0.0204\cdots$ when $g^2=0.405$ and $R_\text{sym}^*-\underline{R}\rightarrow 0$ as $g^2\rightarrow 1$, we get the desired result. 
\end{IEEEproof}

According to Corollary \ref{cor-4}, the HK scheme with Gaussian signals and $|Q|=2$ in Remark \ref{rem-5} is at most $0.125$ bit  away from the capacity for $P\ge 23.3$.

\begin{cor} \label{cor-5} \normalfont
The gap between the TDM lower bound $R_\text{TDM}$ and our upper bound $\overline{R}$ is at most $0.544$ bit in the moderate interference regime. 
\end{cor}

\begin{IEEEproof}
Similar to the proof of Corollary \ref{cor-3}, it suffices to calculate the values of $\overline{R}$ at $g^2=\max(0.086, P^{-1/3}), 0.405, 1$. Using (\ref{eq:NB-20b}) and (\ref{eq:NB-37b}), we can see $\overline{R}-R_\text{TDM} \le 0.544\cdots$, where the equality holds when $g^2=P^{-1/3}=0.086$ (i.e., $P\approx 1584$).
\end{IEEEproof}

The above result implies that the simplest TDM scheme may be sufficient in the moderate interference regime, not requiring rate splitting and simultaneous non-unique decoding. In other words, the interference regime where rate splitting and time sharing become important turns out to be restricted to $g^2\in(0, 0.1)$ and SNR $> 30$ dB except the noisy interference regime.

{Fig. \ref{fig-3a} shows several upper bounds and lower bounds for SNR $=20$ dB.
This figure reveals that the new upper bound in (\ref{eq:NB-20}) is significantly tighter than the known upper bounds over a wide range of the weak interference regime and is lower-bounded by $R_\text{sym}^*$.
Moreover, the rate difference of the simple TDM lower bound from the new upper bound becomes quite small compared to that from the existing upper bounds.


In Fig. \ref{fig-3b}, the same bounds in Fig. \ref{fig-3a} are depicted for even higher SNR of $30$ dB. We can see that $R_\text{sym}^*$ characterizes well the high-SNR capacity of the symmetric GIC in the moderate interference regime. 
The rate gap between the best lower and upper bound depicted by the red dotted curve is now rather negligible in the moderate interference regime at high SNR (i.e., $g^2\in(0.086,1)$ for $P=1000$), which validates Corollary \ref{cor-4}.

\begin{figure}
\hspace{-7mm}
  \includegraphics[scale=.52]{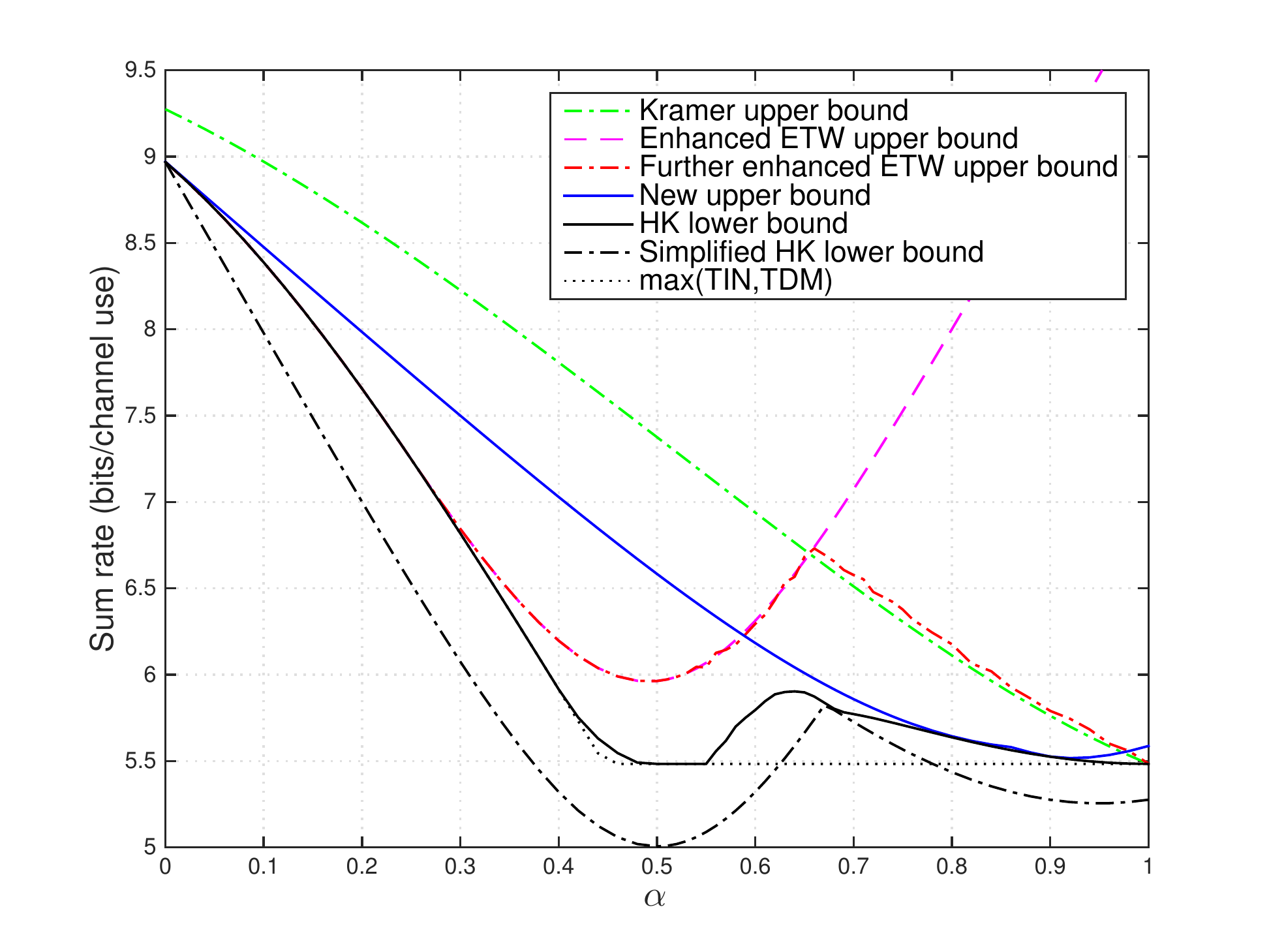}
  \caption{A new look of the ``W curve": The same capacity bounds as Fig. \ref{fig-3b} at SNR $=30$ dB in the different scale of $\alpha=\frac{\log g^2P}{\log P}$.}\label{fig-3c}
\end{figure}

Fig. \ref{fig-3c} shows the same bounds with the same settings of Fig. \ref{fig-3b} in the $\alpha$ scale of the generalized degrees of freedom. We can see that the well-known ``W curve" in \cite[Fig. 11]{Etk08} is not translated well into the behavior of sum-rate bounds in the moderate interference regime even at SNR $=30$ dB. 
This is somehow counter-intuitive to the important insight provided by the {generalized degrees of freedom}  implying that we could obtain a considerable potential gain over the TDM lower bound   around $g^2=P^{-1/3}$ (i.e., $\alpha=2/3$) at high SNR, where the HK scheme was expected to exhibit its promising performance gain. 

According to Figs. \ref{fig-3a} -- \ref{fig-3c}, one might think that there is little point in using a sophisticated interference management scheme in the moderate interference regime. As a consequence, this work points out that the well-known {generalized degrees of freedom} should be carefully interpreted with respect to a real performance gain in the moderate interference regime at practical values of SNR. On the other hand, the remaining regime in which the HK scheme still remains promising is outside the noisy and moderate interference regimes, i.e., the intersection of $|g|(1+g^2P)>0.5$ and $0<g^2<P^{-1/3}$. However, it should be noticed that the values of $g^2$ inside $0<g^2<P^{-1/3}$ (i.e., $0< \alpha< 2/3$) collapse into an arbitrarily small number as $P \rightarrow \infty$.}

\section{High-SNR Capacity Characterization} 
\label{sec:HS}

At this point, we are ready to conduct the high-SNR analysis of the capacity bounds that we have considered so far by using the affine approximation introduced by Shamai and Verd\'u \cite{Sha01}. This is because the generalized degrees of freedom  is only a first-order approximation of the sum capacity, which cannot precisely assess the rate gap between sum-rate lower and upper bounds.  
The high-SNR capacity 􏱖􏰹􏰳$\Cc(P)$ 􏰺has been well approximated by the first-order term (degrees of freedom) and  􏰺the supplementary zero-order term in the expansion of the capacity as an affine function of SNR ($P$) 
\begin{align}  
  \Cc(P) = \Sc (\log P - \Lc) +  o(1) \nonumber 
\end{align}
where 􏱔􏰾$\Sc= \lim_{P\rightarrow \infty} \frac{\Cc(P)}{\log P}$ is the degrees of freedom and $\Lc= \lim_{P\rightarrow \infty} \big(\log P - \frac{\Cc(P)}{\Sc}\big)$ 􏱦􏰾is the power offset in 3-dB units. In the two-user GIC, the degrees of freedom $\Sc$ is one for most transmission strategies and upper-bounding techniques, whereas the power offset can effectively assess the difference among different lower and upper bounds.

The Kramer's upper bound in \cite[Thm. 2]{Kra04} can be rewritten as 
\begin{align} \label{eq:NB-38}
  &R_\text{Kra} = \nonumber \\
  &\hspace{4mm} R_\text{sym}^* +\frac{1}{2}\log \frac{g^2-1+\sqrt{(1+g^2)^2+4g^2(1+g^2)P}}{|g|(1-g^2+\sqrt{(1+g^2)^2+4g^2(1+g^2)P})}.
\end{align}
We denote the power offset of the upper bound ${\min(\overline{R},R_\text{Kra})}$  by $$\Lc_\text{ub} =\lim_{P\rightarrow \infty} \big(\log P - {\min(\overline{R},R_\text{Kra})}\big)$$ and the power offset of the lower bound ${\max(\underline{R}_\text{HK},R_\text{SHK})}$ by $$\Lc_\text{lb} =\lim_{P\rightarrow \infty} \big(\log P - {\max(\underline{R}_\text{HK},R_\text{SHK})}\big).$$ Then, we get the high-SNR rate gap (denoted by $\Delta_\infty$) directly from the difference between the upper and lower bounds on the power offset of $\Cc_\text{sym}(P,g)$ such that $3(\Lc_\text{ub}-\Lc_\text{lb})$  $={\min(\overline{R},R_\text{Kra})}-{\max(\underline{R}_\text{HK},R_\text{SHK})}$ (bps/Hz) $\triangleq \Delta_\infty$, where $3\approx 10\log_{10}2$ is due to the fact that the power offset $\Lc$ is in dB scale by definition (i.e., horizontal offset in capacity versus SNR curves). Using this high-SNR rate gap and Corollary \ref{cor-1}, we have the following result.

\begin{cor} \label{cor-2} \normalfont
The high-SNR rate gap $\Delta_\infty$ between the lower and the upper bounds on the capacity of the symmetric real GIC is at most  
\begin{align} \label{eq:NB-21}
  \Delta_\infty = \left\{ \begin{array}{ll}
  \frac{1}{2}\log \frac{4g^2+1}{2g^2+1} &  \ \phantom{.086}0 < g^2\le 0.086\\
\frac{1}{2}\log \frac{4g^2+1}{4|g|}  &   \ 0.086< g^2\le 0.405\\
\frac{1}{2}\log \frac{2g^2}{\sqrt{4g^2-1}} &  \ 0.405< g^2\le 0.835\\
\frac{1}{2}\log |g|^{-1} &  \ 0.835< g^2\le 1. \\
  \end{array} \right.
\end{align}
In particular, $\Delta_\infty$ vanishes at the values of $g^2=0.25, 0.5$, i.e., for which  
\begin{align} \label{eq:NB-21b}
  \Cc_\text{sym}(P,g) &= R_\text{sym}^* +o(1). 
\end{align} 
\end{cor}

\begin{IEEEproof}
It is straightforward to see from (\ref{eq:NB-38}) and (\ref{eq:NB-36a}) that 
\begin{align} \label{eq:NB-36b}
   R_\text{Kra} &= R_\text{sym}^* +  \frac{1}{2} \log |g|^{-1} +o(1) \nonumber \\
   \underline{R}_\text{HK} &= R_\text{sym}^* +o(1) 
\end{align}
and that $\overline{R}$ and $R_\text{Kra}$ have a crossover point at $g^2= 0.835\cdots$ in the limit of high SNR. 
Then, (\ref{eq:NB-21}) immediately follows from (\ref{eq:NB-24}), (\ref{eq:NB-37}), and (\ref{eq:NB-36b}). The moderate interference regime becomes $0<g^2<1$ as $P^{-1/3} \rightarrow 0$. Moreover, noticing that $
  \frac{1}{2}\log \frac{4g^2+1}{4|g|} = 0 \ \text{ for } g^2=0.25$ and 
$ \frac{1}{2}\log \frac{2g^2}{\sqrt{4g^2-1}} = 0 \ \text{ for } g^2=0.5 $, we have (\ref{eq:NB-21b}). 
\end{IEEEproof}

\begin{figure*}
\center
  \includegraphics[scale=.75]{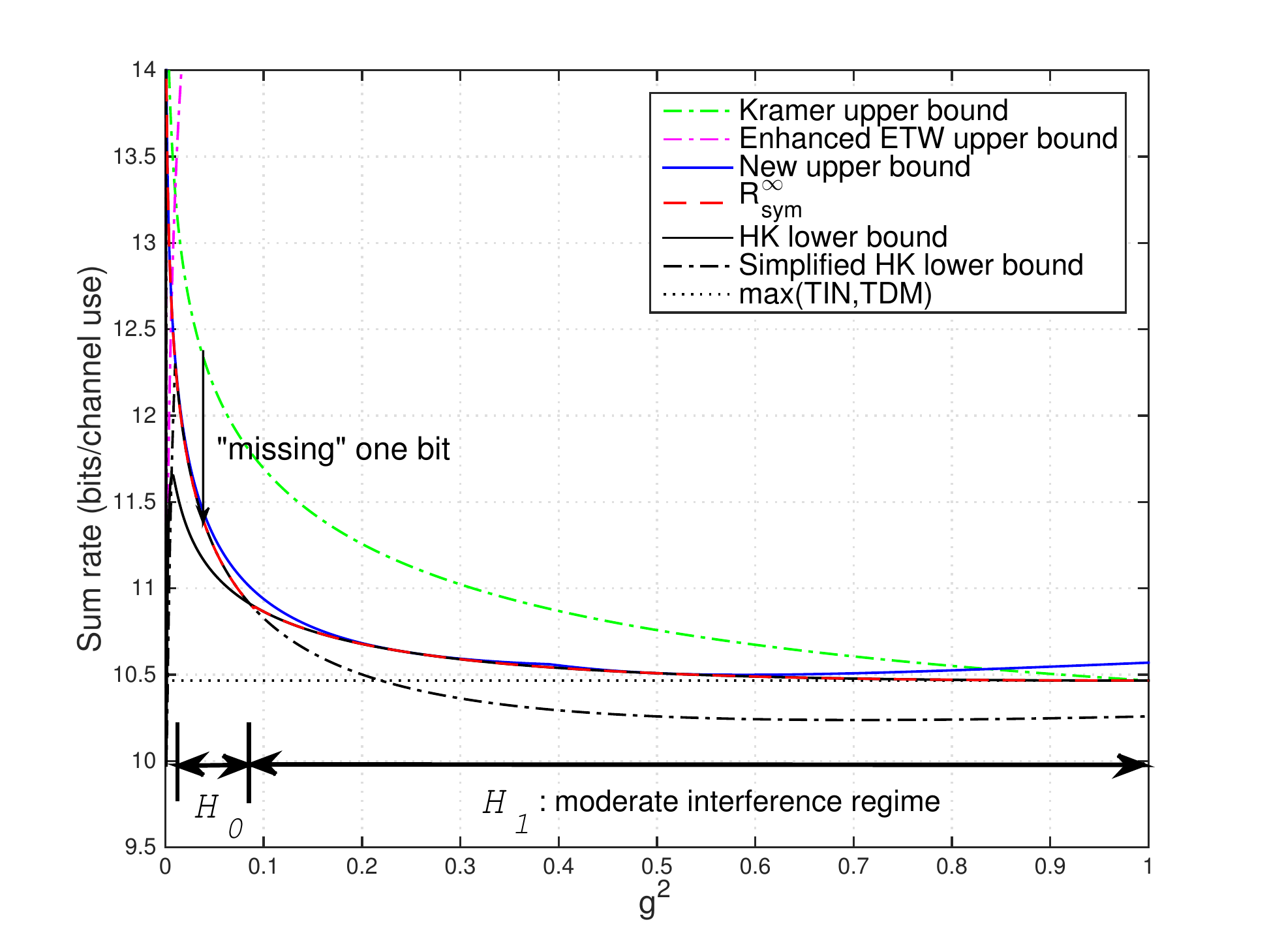}
  \caption{Bounds on the sum capacity of the symmetric GIC at SNR $=60$ dB, where $R_\text{sym}^\infty$ is given by (\ref{eq:intro-1b}).}\label{fig-3d}
\end{figure*}

Finally, it would deserve to mention 
\begin{align} 
  \Delta_\infty = \left\{ \begin{array}{ll}
  0.098 &  \ g^2 = 0.086\\
  0.021 &  \ g^2= 0.405\\
  0.063 &  \ g^2= 0.835
  \end{array} \right. \nonumber
\end{align}
and $\Delta_\infty$ has its largest value at $g^2 = 0.086$. Thus, the high-SNR rate gap is at most 0.098 ($\approx$ 0.1) bit per channel use for $0< g^2 < 1$. 

{
It is well known that the generalized degrees of freedom is $1-\frac{\alpha}{2}$ for $\frac{2}{3}\le \alpha \le 1$ \cite{Etk08}. Based on our results, we can identify that there are two subregimes in $\frac{2}{3}\le \alpha \le 1$ (i.e., $P^{-\frac{1}{3}}\le g^2 \le 1$) at high SNR. Say, $H_0 = \{g^2: P^{-\frac{1}{3}}\le g^2 < 0.086\}$ and $H_1 = \{g^2: 0.086 \le g^2\le 1\}$. Let $R_\text{sym}^\infty$ denote the high-SNR characterization of the capacity of the symmetric real GIC. For the subregimes, we have the following result.

\begin{thm} \label{cor-4b} \normalfont
For $P^{-\frac{1}{3}}\le g^2\le 1$, the high-SNR capacity can be characterized to within $0.1$-bit gap by
\begin{align} \label{eq:intro-1b}
    R_\text{sym}^\infty  = \left\{ \begin{array}{ll}
  R_\text{sym}^* +\frac{1}{2}\log (2|g|+|g|^{-1})-1 & g^2\in H_0 \\ 
  R_\text{sym}^* &  g^2\in H_1 .
  \end{array} \right. 
\end{align}
In particular, the high-SNR ratio\footnote{The ratio $r$ here does \emph{not necessarily} stipulate $P\rightarrow \infty$ with some abuse of notation unlike the (generalized) degrees of freedom or $\Delta_\infty$. This is because when one converts the scale of $g^2$ to that of $\alpha$, all values of $g^2 \in (0,1]$ collapse into $\alpha=1$ by definition as $P\rightarrow\infty$.} $r\triangleq \frac{R_\text{sym}^\infty}{\frac{1}{2}\log \snr}$ is given by
\begin{align} \label{eq:intro-1c}
    r  \approx \left\{ \begin{array}{ll}
  1-\frac{\alpha}{2} & \ g^2\in H_0  \\
  \frac{3-\alpha}{4} & \ g^2\in H_1  .
  \end{array} \right. 
\end{align}
\end{thm}

\begin{IEEEproof}
It suffices to check $R_\text{sym}^\infty$ for $g^2\in H_0$. Comparing $\overline{R}$ in (\ref{eq:NB-24}) and $R_\text{SHK}$ in (\ref{eq:NB-37}), we can see that $\frac{1}{2}\log \frac{4g^2+1}{4|g|} \approx \frac{1}{2} \log \frac{2|g|+|g|^{-1}}{4} =\frac{1}{2}\log (2|g|+|g|^{-1})-1$ at high SNR for small $g^2$. Moreover, it follows from Corollary \ref{cor-2} that the high-SNR capacity characterization $R_\text{sym}^\infty$ is at most 0.1 bit far from the capacity for $P^{-\frac{1}{3}}\le g^2\le 1$. The ratio $r$ immediately follows from substituting $|g|=P^{\frac{\alpha-1}{2}}$ into (\ref{eq:intro-1b}) by the definition of $\alpha$. 
\end{IEEEproof}

In fact, $R_\text{sym}^\infty =\max(R_\text{SHK}, \underline{R})$ at high SNR. 
The ratio $r$ can be used to predict the high-SNR performance gain of the HK scheme relative to the simple TDM scheme whose ratio is $\frac{R_\text{TDM}}{\frac{1}{2}\log \snr}\approx 1/2$. 
Notice that $\frac{1}{2}\log (2|g|+|g|^{-1})=1$ at $|g|=1-\sqrt{1/2}\approx \sqrt{0.086}$. This means that $g^2=0.086$ in the moderate interference regime in Definition \ref{def-1} is also the point where $\underline{R}$ in (\ref{eq:NB-36b}) and $R_\text{SHK}$ coincide at high SNR.  It is clear that the above high-SNR characterization holds true for the entire regime of interest, i.e., $0<g^2\le 1$, as $P\rightarrow\infty$. 

\begin{rem} \label{rem-2} \normalfont
It follows from Corollary \ref{cor-2} and a close look at (\ref{eq:NB-36a}) that the time-sharing operation on messages $\mathcal{U}_i$ and $\mathcal{W}_i$ in Remark \ref{rem-5} yields $\frac{1}{2}\log|g|^{-1}$-bit gain at high SNR, compared to the simple TDM scheme. This time sharing together with rate splitting allows only a \emph{single} virtual MAC at either of the receivers at a certain time. For the converse, the time sharing on genie signals of the proposed upper bound in (\ref{eq:IS-2}) does not increase the high-SNR capacity at least at $g^2=0.25, 0.5$ ($\frac{1}{2}\log|g|^{-1}$ appears due to Step $A3)$ in (\ref{eq:NB-25})). Therefore, the time sharing gain seems at most $\frac{1}{2}\log|g|^{-1}$ bit at high SNR in the moderate interference regime ($g^2\in H_1$). On the other hand, the simplified HK scheme benefits from creating \emph{two} virtual MACs by rate splitting without time sharing. It follows from \eqref{eq:intro-1b} that the resulting $\log|g|^{-1}$-bit gain of such two virtual MACs is double the gain of the single virtual MAC in (\ref{eq:NB-36a}). On the other hand, the performance gain of the simplified HK scheme is significant only when $g^2\in H_0$ and SNR is sufficiently high ($> 30$ dB).  
Again for the converse, the proposed time sharing on genie signals in this work does not increase the high-SNR capacity at $g^2=P^{-\frac{1}{3}}\rightarrow 0$.
\end{rem}

}

{Fig. \ref{fig-3d} validates the two effective subregimes $H_0$ and $H_1$ with different ratios $r$ in Theorem \ref{cor-4b} at very high SNR of $60$ dB. It is also shown that the ``missing" one bit per channel use is \emph{almost} found in $H_0$. Namely, the proposed upper bound is at most 0.1 bit far from the HK bounds and also asymptotically tight at the values of $g^2=P^{-\frac{1}{3}}\rightarrow 0$.}



\section{New Outer Bounds on the Capacity Region}
\label{sec:OB}

In Sec. \ref{sec:NB}, we have shown that the sum-rate upper bound can be significantly tightened by jointly utilizing the change-of-interference approach in conjunction with the Etkin-type approach. In this section, we use the same technique to improve outer bounds on the capacity region of the two-user GIC. As before, we restrict our attention only to the following combinations: 
\begin{align} 
  n(2R_1+R_2-3\epsilon_n) &\le I(X_{1}^n;Y_{1}^n|X_{2}^n)+I(X_{1}^n;Y_{1}^n) \nonumber \\ &\ \ \ +I({X}_{2}^n;{Y}_{2}^n|U_{2}^n) \label{eq:ob-7f} \\ 
  n(R_1+2R_2-3\epsilon_n) &\le I({X}_{1}^n;{Y}_{1}^n|U_{1}^n)+I(X_{2}^n;Y_{2}^n|X_{1}^n) \nonumber \\ &\ \ \ +I(X_{2}^n;Y_{2}^n) \label{eq:ob-7g} \\
  n(R_1+2R_2-2\epsilon_n) &\le I({X}_{1}^n;{Y}_{1}^n,S_{1}^n) +I({X}_{2}^n;{Y}_{2}^n|U_{2}^n) \nonumber \\ &\ \ \  +I({X}_{2}^n;{Y}_{2}^n) \label{eq:ob-7a} \\
  n(2R_1+R_2-2\epsilon_n) &\le I({X}_{1}^n;{Y}_{1}^n|U_{1}^n) +I(X_{1}^n;Y_{1}^n) \nonumber \\ &\ \ \ +I({X}_{2}^n;{Y}_{2}^n,S_{2}^n). \label{eq:ob-7b} 
\end{align}

Using a known bounding technique by Kramer {\cite[Thm. 2]{Kra04}}, we can obtain upper bounds for both $R_1+2R_2$ in (\ref{eq:ob-7f}) and $2R_1+R_2$ in (\ref{eq:ob-7g}) as follows. 

\begin{thm} \label{thm-2} \normalfont
The capacity region of the two-user GIC  in the weak interference regime is contained in the following region: 
\begin{align} 
  &R_1\le h(U_{1G}) -h(h_{12}Y_{2G}+ \tilde{V}_{W_1}|U_{2G}) \nonumber \\
  &\; +h(Y_{1G}|U_{1G}) -h(W_1-Z_1) \nonumber \\ 
  &\; -h(h_{21}Y_{1G}+ \tilde{V}_{W_2}|U_{1G})+h(Y_{2G}|U_{2G})  -h(W_2-Z_2)  \nonumber \\ &\; +\frac{1}{2}\log \Big(2\pi e\big((P_2+|h_{21}|^2P_1+1)2^{-2R_2}-1+\sigma_{W_2}^2 \big) \Big) -R_2 \label{eq:B-11a} 
\end{align}
for all $(W_1, W_2)$ satisfying (\ref{eq:B-8b}), (\ref{eq:B-8a}), and  $\sigma_{W_2}^2\le 1$, where $\tilde{V}_{W_1}$ is defined in (\ref{eq:B-8c}).
\begin{align} 
  &R_2\le h(U_{2G}) -h(h_{21}Y_{1G}+ \tilde{V}_{W_2}|U_{1G}) \nonumber \\
  &\; +h(Y_{2G}|U_{2G})-h(W_2-Z_2) \nonumber \\ 
  &\; -h(h_{12}Y_{2G}+ \tilde{V}_{W_1}|U_{2G})+h(Y_{1G}|U_{1G}) -h(W_1-Z_1) \nonumber \\ 
  &\; +\frac{1}{2}\log \Big(2\pi e\big((P_1+|h_{12}|^2P_2+1)2^{-2R_1}-1+\sigma_{W_1}^2 \big) \Big) -R_1 \label{eq:B-11b}
\end{align}
for all $(W_1, W_2)$ satisfying (\ref{eq:B-8b}), (\ref{eq:B-8a}), and  $\sigma_{W_1}^2\le 1$, where $\tilde{V}_{W_2}$ is defined in (\ref{eq:B-8d}).  Interchanging the user indices, we can have another bound for $2R_1+R_2$. 
\end{thm}

\begin{IEEEproof}
Similar to the proof of Theorem \ref{thm-1}, using Lemma \ref{lem-8}, we can first bound $R_1+R_2$ as 
\begin{align} \label{eq:B-12}
  R_1&+R_2\le h(U_{1G}) -h(h_{12}Y_{2G}+ \tilde{V}_{W_1}|U_{2G})+h(Y_{1G}|U_{1G}) \nonumber \\ &\ \ \ -h(W_1-Z_1)  +\frac{1}{n}h(U_{2}^n) -h(h_{21}Y_{1G}+ \tilde{V}_{W_2}|U_{1G}) \nonumber \\ &\ \ \ +h(Y_{2G}|U_{2G})-h(W_2-Z_2).
\end{align}
Following the EPI bounding technique in \cite[Lem. 11]{Ann09} (see also \cite[Thm. 2]{Kra04}) based on the known outer bound results for the degraded GIC \cite{Sat78} and the one-sided GIC \cite{Cos85}, we have 
\begin{align} \label{eq:B-15}
  h(&U^n_2)\nonumber \\ 
  &= h(h_{21}X^n_1+W_2^n) \nonumber \\ 
  &\overset{(a)}{\ge} \frac{n}{2} \log \Big(2\pi e\big(2^{\frac{2}{n}h(h_{21}X^n_1+Z_2^n)}-(1-\sigma_{W_2}^2)\big)\Big) \nonumber \\ 
  &\overset{(b)}{\ge} \frac{n}{2}\log \Big(2\pi e\big((P_2+|h_{21}|^2P_1+1)2^{-2R_2}-1+\sigma_{W_2}^2 \big) \Big) 
\end{align}
where $(a)$ follows from EPI and $(b)$ follows from (\ref{eq:B-22}). 
Substituting (\ref{eq:B-15}) into (\ref{eq:B-12}) gives the upper bound (\ref{eq:B-11a}) on $R_1+2R_2$. Interchanging the user indices, we have (\ref{eq:B-11b}) for $2R_1+R_2$. 
\end{IEEEproof}

Using (\ref{eq:ob-7a}) and (\ref{eq:ob-7b}), we derive the following result to upper-bound $R_1+2R_2$ and $2R_1+R_2$.

\begin{thm} \label{thm-0} \normalfont
For all parameters in (\ref{eq:NB-16}) such that 
\begin{align} 
  \boldsymbol{\kappa} \in \{\boldsymbol{\kappa}: \sigma_{V_{W_2}}^2\ge \min(\sigma_{N_1}^2, \sigma_{W_2}^2),\; \sigma_{V_{N_1}}^2\ge |h_{12}|^2\sigma_{Z_2-W_2}^2\} \label{eq:B-18b} 
\end{align}
the capacity region of the two-user GIC  in the weak interference regime is contained in the following region: 
\begin{align} \label{eq:NB-4}
  R_1+2R_2 &\le h(S_{1G}) -h(h_{21}X_{1G}+V_{W_2})  \nonumber \\ 
  &+h(Y_{2G}|U_{2G}) -h(h_{12}Y_{2G}+ \tilde{V}'_{N_1}|U_{2G}) \nonumber \\ 
  &+h(Y_{1G}|S_{1G}) -h(N_1) +h(U_{2G}) -h(Z_2-W_2).  
\end{align} 
 Interchanging the user indices, we can have another bound for $2R_1+R_2$. 
\end{thm}

\begin{figure*}
\hspace{-7mm}
\center
  \includegraphics[scale=.75]{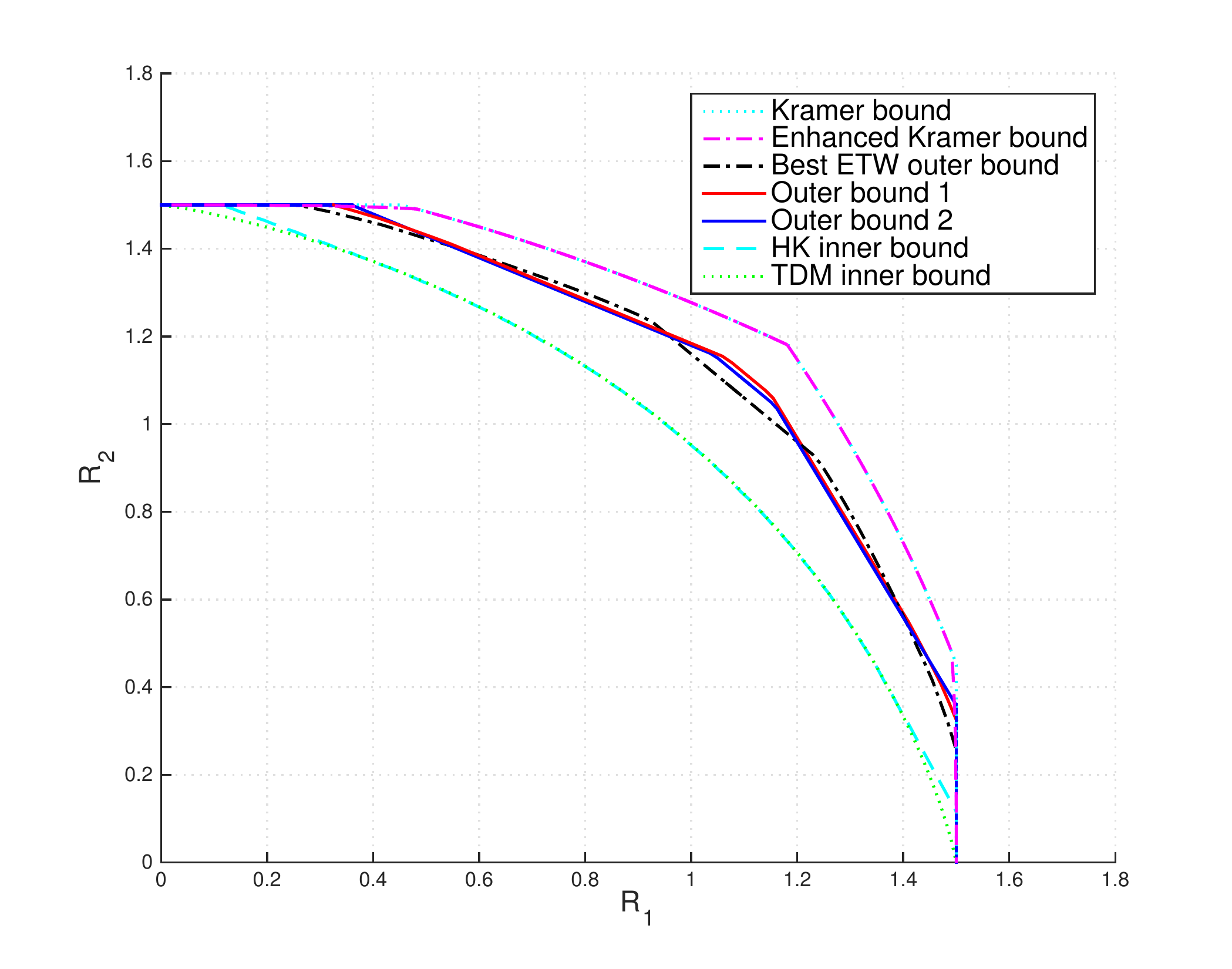}
  \caption{Bounds on the capacity region of two-user symmetric Gaussian interference channel: $P=7$ and $g^2=0.2$. {The HK inner bound is given by \cite[Eq. (18)]{Sas04}}}\label{fig-5}
\end{figure*}

\begin{IEEEproof}
Using (\ref{eq:BT-4b}), (\ref{eq:B-22}), and (\ref{eq:GO-1}), we can bound $R_1+2R_2$ as follows.
\begin{subequations} 
\begin{align} 
  n&(R_1+2R_2-2\epsilon_n) \nonumber \\ 
  &\le h(S^n_1)- nh(N_1) +h(Y^n_1|S^n_1)-h(h_{12}X^n_2+V^n_{N_1})  \nonumber \\ 
  & \ \ \ +h(Y^n_2|U^n_2) - h(h_{21}X^n_1+V^n_{W_2}) +h(U^n_2) \nonumber \\ 
  &\ \ \ -nh(Z_2-W_2) +nh(Y_{2G})-h(Y^n_2|X^n_2) \nonumber \\ 
  &= h(S^n_1) -h(h_{21}X^n_1+V^n_{W_2}) +h(U^n_2) -h(Y^n_2|X^n_2) \label{eq:NB-2a} \\ 
  & \ \ \ +h(Y^n_2|U^n_2) -h(h_{12}X^n_2+V^n_{N_1})  \label{eq:NB-2b} \\ 
  & \ \ \ +h(Y^n_1|S^n_1) -nh(N_1) +nh(Y_{2G})-nh(Z_2-W_2).  \label{eq:NB-2c} 
\end{align}
\end{subequations} 
The upper bound of (\ref{eq:NB-2a}) is given in (\ref{eq:NB-3}).  Also if (\ref{eq:B-18b}) holds, (\ref{eq:NB-2b}) can be bounded by Lemma \ref{lem-8} as
\begin{align}  \label{eq:B-7b}
  h(Y^n_2|U^n_2)&-h(h_{12}X^n_2+V^n_{N_1}) \le  \nonumber \\ 
  &nh(Y_{2G}|U_{2G}) -nh(h_{12}Y_{2G}+ \tilde{V}_{N_1}|U_{2G}). 
\end{align}
Finally applying the argument that Gaussian maximizes entropy to (\ref{eq:NB-2c}) and using (\ref{eq:NB-3}) and (\ref{eq:B-7b}), we obtain (\ref{eq:NB-4}).
\end{IEEEproof}

As before, we can rewrite (\ref{eq:NB-4}) as
\begin{align} \label{eq:NB-10c}
  R_1\;+\;&2R_2 \nonumber \\ 
  &\le \log \left(\frac{|h_{21}|^2P_1+\sigma_{N_1}^2}{|h_{21}|^2P_1+\sigma_{V_{W_2}}^2}\right) +\log \left(\frac{|h_{21}|^2P_1+\sigma_{W_2}^2}{|h_{21}|^2P_1+1}\right) \nonumber \\ &\ 
+\log \left(\frac{P_2+|h_{21}|^2P_1+1-\frac{(|h_{21}|^2P_1+\rho_{W_2}\sigma_{W_2})^2}{|h_{21}|^2P_1+\sigma_{W_2}^2}}{|h_{12}|^2P_2+\sigma_{V_{N_1}}^2-\frac{(h_{12}\rho_{W_2}\sigma_{W_2}-h_{12}\sigma_{W_2}^2)^2}{|h_{21}|^2P_1+\sigma_{W_2}^2}}\right) 
\nonumber \\ &\ 
+ \log \left(\frac{P_1+|h_{12}|^2P_2+1-\frac{(h_{21}P_1+\rho_{N_1}\sigma_{N_1})^2}{|h_{21}|^2P_1+\sigma_{N_1}^2}}{\sigma_{N_1}^2}\right)  \nonumber \\ 
&\ +\log \left(\frac{P_2+|h_{21}|^2P_1+1}{1+\sigma_{W_2}^2-2\rho_{W_2}\sigma_{W_2}}\right) .
\end{align}

Fig. \ref{fig-5} depicts the new outer bounds on the capacity region of the two-user symmetric GIC for $P=7$ and $g^2=0.2$, where the best ETW outer bound is given by the intersection of \cite{Mot09,Ann09} and \cite{Etk09}. While  outer bound 1 is the intersection of Theorems \ref{thm-1} and \ref{thm-2}, outer bound 2 is taken from the intersection of Theorems \ref{thm-6} and \ref{thm-0}. We also plot the time division inner bound. Notice that we can obtain a more sophisticated inner bound by using a simplified case of the HK inner bound that does not consider time sharing and is limited to only Gaussian distributions for the private and common messages. Our capacity outer bounds are shown to be tighter than the best genie-aided bound for a certain region of the rate pair $(R_1,R_2)$. 

\begin{figure*}
\hspace{-6mm}
\center
  \includegraphics[scale=.75]{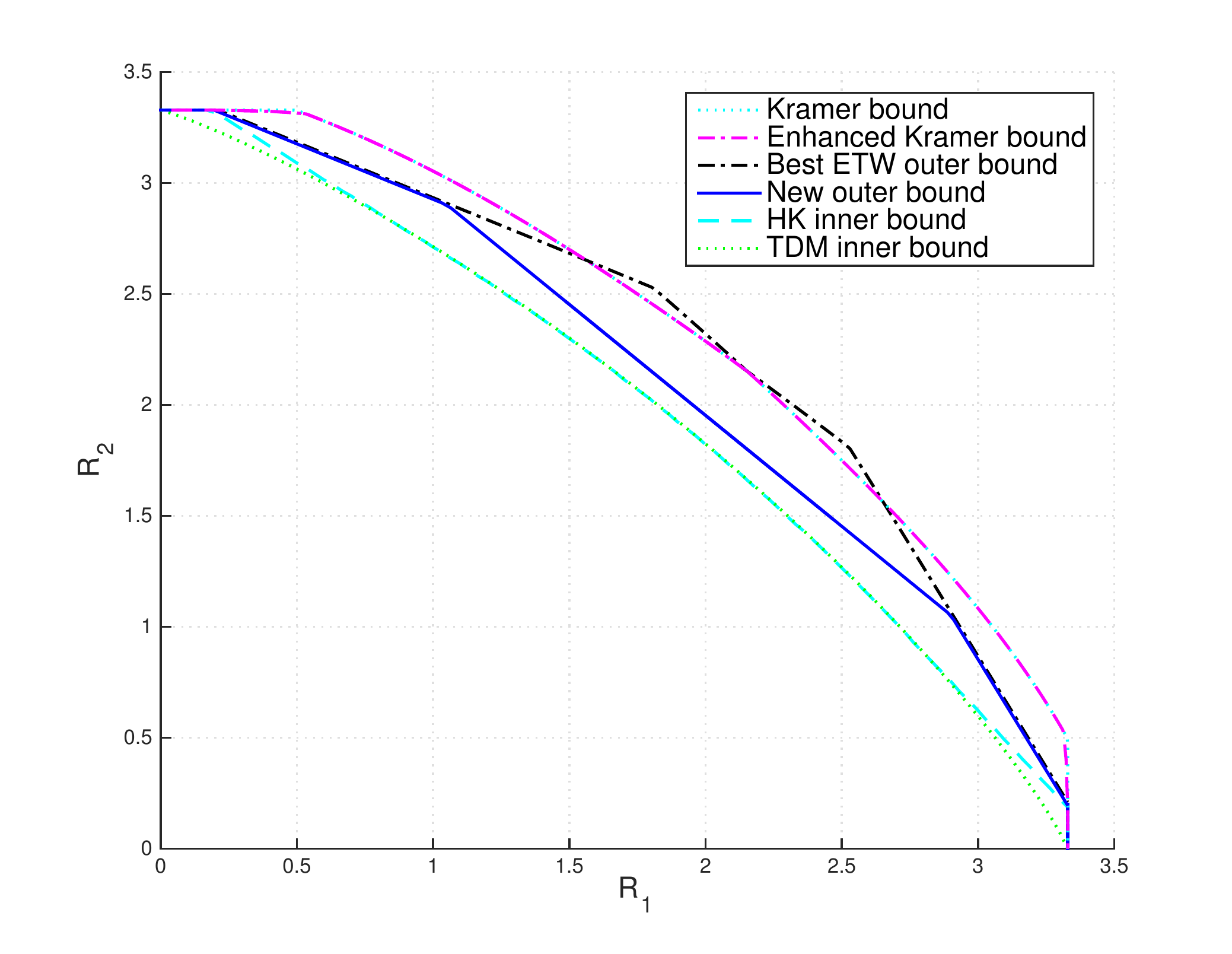}
  \caption{Bounds on the capacity region of two-user symmetric Gaussian interference channel: $P=100$ (SNR $=20$ dB) and $g^2=0.3$.}\label{fig-6}
\end{figure*}

Fig. \ref{fig-6} illustrates the capacity outer bounds for $P=100$ and $g^2=0.3$. The new outer bound indicates outer bound 2, whereas outer bound 1 is relatively quite loose and hence omitted here. In this case, the new outer bound is not outperformed by the best genie-aided bound.
{The enhanced Kramer outer bound is shown to be only tightest around the corner points.  Although the bound settles the ``missing corner point" conjecture, the corresponding outer-bound enhancement seems limited. 
}

\section{Conclusion}
\label{sec:Con}

In this paper, we have developed a new genie-aided approach referred to as change-of-interference that makes the outer-bounding problem for the two-user GIC mathematically tractable by changing  arbitrarily distributed interference into a Gaussian random sequence. In order to upper-bound the resulting differential entropy terms, the conditional worst additive noise lemma was introduced. We have also identified some useful combinations of the change-of-interference approach with the genie-aided approach to get  tighter outer bounds in the weak interference regime. 
{A main result is that we have characterized the sum capacity to within $0.104$ bit in the moderate interference regime and showed the high-SNR capacity at certain points of channel coefficients.  Hence, the proposed genie-aided approach proved itself useful to better understand the two-user GIC.}
Another implication is that any sophisticated interference management scheme cannot achieve a significant performance gain over the simple time division scheme  in the moderate interference regime for the symmetric real GIC. 
Namely, the simplest TDM scheme may be sufficient in the moderate interference regime. 
Therefore, this work points out that the well-known generalized degrees of freedom result should be carefully interpreted into a real performance gain in the moderate interference regime at practical values of SNR. 

{The known capacity upper bounds \emph{without time sharing} on side information at the receivers are far from the HK lower bound as large as $1$ bit per channel use especially at high SNR. We have proposed the upper-bounding strategy that exploits time sharing on side information (instead of channel inputs or messages). Namely, only with probability $1/2$ a genie provides either of the receivers with its noisy input signal and the other receiver with its noisy interference. We have shown that the HK achievable scheme approaches the proposed upper bound with at most 0.1 bit gap at high SNR. Therefore, it is argued that the notion of \emph{time sharing} is not just essential in the HK scheme but also fundamental in characterizing the sum capacity (or finding the ``missing" one bit) at high SNR. Moreover, we have identified the two subregimes in the weak interference regime at high SNR, which could not be revealed with the generalized degrees of freedom.}

We conclude this paper by pointing out some possible extensions of the ideas presented in this work for future work. First, there may be other useful combinations or more general change-of-interference signals in Appendix \ref{sec:app-0} that lend themselves to tighter outer bounds than the proposed bounds in this work. A prospective study is to see if $R_\text{sym}^*$ is indeed the high-SNR sum capacity for all non-zero channel coefficients. Second, the new bounding techniques can be used for the more than two users GIC, for instance, which can be found in our companion work \cite{Nam15b}, and possibly even for other wireless networks with mutually interfering links. Lastly, it is of interest if we can extend our capacity characterization for the symmetric real GIC to the case of the general \emph{asymmetric complex} GIC.

\section*{Acknowledgement} 
The author would like to thank Giuseppe Caire for his comment on the preliminary version of this work and for his consistent support. He also thanks Daniela Tuninetti, the Associate Editor, for her valuable feedback that improved the work.

\appendices

\section{General Form of Change-of-Interference Approach}
\label{sec:app-0}

We begin with introducing additional random sequences $T_i^n$ that are independent of $U_i$ for $i=1,2$.
\begin{align} 
    n(R_1-\epsilon_n) &\le I(X^n_1;Y^n_1) \nonumber \\ 
  &=  h(Y^n_1)-h(Y^n_1|X^n_1)  \nonumber \\ 
  &= I(U^n_1;Y^n_1)+h(Y^n_1|U^n_1)-h(Y^n_1|X^n_1)  \nonumber \\ 
  &\le I(U^n_1;Y^n_1|T^n_1)+h(Y^n_1|U^n_1)-h(Y^n_1|X^n_1) \label{eq:BT-2}\\
  &= h(U^n_1)-h(U^n_1|Y^n_1,T^n_1)+h(Y^n_1|U^n_1) \nonumber \\
  &\hspace{5mm} -h(Y^n_1|X^n_1) \label{eq:BT-2b}
 \end{align}
where \eqref{eq:BT-2} follows from the independence assumption between $U_1^n$ and $T_1^n$ and from (\ref{eq:BT-8}). 
Letting 
\begin{align} \label{eq:A-4a}
  T_1=X_1 \ \text{ and } \ T_2=X_2
\end{align}
(\ref{eq:BT-2}) reduces to the special case, $I(X^n_1;Y^n_1|U^n_1)$ {in (\ref{eq:BT-4a})}. We can define a more general $T_1^n$ such that 
\begin{align} \label{eq:A-4}
  T_1&=X_1+W'_1 \nonumber \\ T_2&=X_2+W'_2
\end{align}
where $W'_i \sim \mathcal{CN}(0,\sigma_{W'_i}^2)$ is possibly correlated to $Z_i$ for $i=1,2$, but independent of everything else. Then, the second term in (\ref{eq:BT-2b}) can be lower-bounded as
\begin{align} \label{eq:BT-6}
  h&(U^n_1|Y^n_1,T^n_1) \nonumber \\
  &= h(h_{12}X^n_2+W^n_1|X_1^n+h_{12}X^n_2+Z^n_1, X_1^n+{W'}_1^n)  \nonumber \\
  &\ge  h(h_{12}X^n_2+W^n_1|X_1^n+h_{12}X^n_2+Z^n_1, X_1^n+{W'}_1^n,X_1^n) \nonumber \\
  &\overset{(a)}{=}  h(h_{12}X^n_2+W^n_1|h_{12}X^n_2+Z^n_1)   \nonumber \\
  &=  h(U^n_1|Y^n_1,X^n_1) 
\end{align}
where $(a)$ follows from the fact that $Y_1^n \;\rightarrow\; X_1^n\;\rightarrow \;X_1^n+{W'}_1^n$ forms a Markov chain by the assumption on ${W'}_1^n$.  This shows that $I(U^n_1;Y^n_1|T^n_1)\le I(U^n_1;Y^n_1|X^n_1)$. Therefore, the choice of (\ref{eq:A-4}) may further improve the upper bound in (\ref{eq:BT-5b}).

\section{Proof of Lemma \ref{lem-4} and Corollary \ref{lem-2}}
\label{sec:app-1}

Without loss of generality, we assume that $\frac{1}{n}\sum_{j=1}^n\mathrm{Cov}(Z_j,X_j+Z_j,U_j)= \pK$. Following the proof in \cite[Lem. 4]{Ann09}, we have
\begin{align} 
  h(X^n|U^n)-&h(X^n+Z^n|U^n)  \nonumber \\
  &=-I(Z^n;X^n+Z^n|U^n) \nonumber \\
  &= -h(Z^n|U^n)+h(Z^n|X^n+Z^n,U^n) \nonumber \\
  &\overset{(a)}{\le} -nh(Z|U_g)+nh(Z|X_g+Z,U_g) \nonumber \\
  &= nh(X_g|U_g)-nh(X_g+Z|U_g) 
\end{align}
where $(a)$ follows from the facts that $Z^n$ is independent of $U^n$ and {that Gaussian maximizes entropy}.

The following is a somewhat straightforward generalization of the worst additive noise lemma in \cite{Dig01} to be used in this work.

\begin{cor}\label{lem-2} \normalfont
Let $X^n$ denote a random sequence with an average power constraint, $W^n$ and $Z^n$ be i.i.d. $\mathcal{N}(0,\sigma_W^2)$ and $\mathcal{N}(0,\sigma_Z^2)$, respectively, correlated with each other but independent of $X^n$. If $\sigma_Z^2-\sigma_W^2\ge 0$, then
\begin{align} \label{eq:B-23}
  h(X^n+W^n)-&h(X^n+Z^n) \nonumber \\
  &\le nh(X_g+W)-nh(X_g+Z)
\end{align}
where the equality holds if $X^n=X^n_g$.
\end{cor}

\begin{IEEEproof} 
We only present a sketch of the proof.  Let $V^n$ be i.i.d. $\mathcal{N}(0,\sigma_Z^2-\sigma_W^2)$ independent of all other random sequences since $\sigma_Z^2-\sigma_W^2\ge 0$, then (\ref{eq:B-23}) immediately follows from the proof of \cite[Lem. 4]{Ann09}. 
\end{IEEEproof}

\section{Proof of Lemma \ref{lem-8}}
\label{sec:app-2}

We can rewrite the left-hand side of (\ref{eq:B-20}) as
\begin{align} 
  h(&X^n+Y^n+Z^n|Y^n+W^n) - h(X^n+V^n) \nonumber \\
  &= h(X^n+Z^n-W^n|Y^n+W^n)- h(X^n+V^n) \nonumber \\
  &\overset{(a)}{=} h(X^n+Z^n-W^n|Y^n+W^n) \nonumber \\
  &\hspace{5mm} - h(X^n+Z^n-W^n+\sqrt{1-\sigma_{V}^{-2}\sigma_{Z- W}^2}\;V^n) \nonumber \\
  &\overset{(b)}{\le} h(X^n+Z^n-W^n|Y^n+W^n) \nonumber \\
  &\hspace{5mm} - h(X^n+Z^n-W^n+\tilde{V}^n|Y^n+W^n)\nonumber \\
  &\overset{(c)}{\le} nh(X_g+Y_g+Z|Y_g+W) \nonumber \\
  &\hspace{5mm} - nh(X_g+Y_g+Z+\tilde{V}|Y_g+W) \nonumber \\
  &= nh(X_g+Y_g+Z|Y_g+W) \nonumber \\
  &\hspace{5mm} - nh(X_g+Y_g+Z+\tilde{V}|Y_g+W) \nonumber
\end{align}
where $(a)$ follows from the facts that the i.i.d. Gaussian random sequences {$V^n$ and $Z^n-W^n+\tilde{V}^n$ are statistically equivalent} and that the condition in (\ref{eq:B-24}) satisfies $\sigma_{\tilde{V}}^{2}=\sigma_{V}^{2}-\sigma_{Z- W}^2\ge 0$, $(b)$  follows from the fact that conditioning reduces entropy, and $(c)$ follows by using the fact that $\tilde{V}^n$ is independent of $W^n$ and by applying Lemma \ref{lem-4} for an average covariance constraint on the random vector sequence $(\tilde{V},X+Y+Z+\tilde{V},Y+W)^n$.

\bibliographystyle{IEEEtran}
\bibliography{OB_2user_GIC}

\begin{IEEEbiographynophoto}{Junyoung Nam}  (M'17)
received the B.Sc. in Statistics from Inha University, Incheon, Korea, in 1997 and the M.Sc. and the Ph.D. degrees in Electrical Engineering (Information and Communication) from Korea Advanced Institute of Science and Technology (KAIST), Daejeon, Korea, in 2008 and 2015, respectively.    

Dr. Nam was with the Communications R\&D Center, Samsung Electronics, Seoul, Korea, from 1997 to 2001, and also with the Communications Lab., Seodu InChip, Seoul, from 2001 to 2006. In Summer 2006, he joined Electronics and Telecommunications Research Institute (ETRI), Deajeon, Korea, where he was a principal member of research staff. Since 2017, he has been with the department of Wireless Communications and Networks, Fraunhofer Heinrich Hertz Institute (HHI), Berlin, Germany. His research interests are in the areas of wireless communications, information theory, cellular system design, and compressed sensing.  
\end{IEEEbiographynophoto}

\end{document}